\documentclass{aastex61}
 \usepackage{booktabs}
 \usepackage{multirow}
 \usepackage{hyperref}
 \usepackage{cleveref}
 \usepackage{ifpdf}
 \ifpdf
 \else
 \fi

\newcommand\aastex{AAS\TeX}

\received{}
\revised{}
\accepted{\today}
\submitjournal{}

\shorttitle{\aastex\ LSBG catalogue and Tully-Fisher relation}
\shortauthors{Du et al.}

\begin{document}

\title{Low Surface Brightness Galaxy catalogue selected from the $\alpha$.40-SDSS DR7 Survey and Tully-Fisher relation}

\correspondingauthor{Du Wei}
\email{wdu@nao.cas.cn}

\author{Du Wei}
\affil{National Astronomical Observatories, Chinese Academy of Sciences, 20A Datun Road, Chaoyang District, Beijing 100012, China}
\affil{Key Laboratory of Optical Astronomy, National Astronomical Observatories, Chinese Academy of Sciences, 20A Datun Road, Chaoyang District, Beijing 100012, China}

\author{Cheng Cheng}
\affil{National Astronomical Observatories, Chinese Academy of Sciences, 20A Datun Road, Chaoyang District, Beijing 100012, China}
\affiliation{Chinese Academy of Sciences South America Center for Astronomy, National Astronomical Observatories, Chinese Academy of Sciences, 20A Datun Road, Chaoyang District, Beijing 100012, China}
\nocollaboration

\author{Wu Hong}
\affil{National Astronomical Observatories, Chinese Academy of Sciences, 20A Datun Road, Chaoyang District, Beijing 100012, China}
\affiliation{Key Laboratory of Optical Astronomy, National Astronomical Observatories, Chinese Academy of Sciences, 20A Datun Road, Chaoyang District, Beijing 100012, China}
\nocollaboration

\author{Zhu Ming}
\affil{National Astronomical Observatories, Chinese Academy of Sciences, 20A Datun Road, Chaoyang District, Beijing 100012, China}
\affiliation{Chinese Academy of Sciences Key Laboratory of FAST, National Astronomical Observatories, Chinese Academy of Sciences, 20A Datun Road, Chaoyang District, Beijing 100012, China}
\nocollaboration

\author{Wang Yougang}
\affil{National Astronomical Observatories, Chinese Academy of Sciences, 20A Datun Road, Chaoyang District, Beijing 100012, China}
\affiliation{Key Laboratory of Computational Astrophysics, National Astronomical Observatories, Chinese Academy of Sciences, 20A Datun Road, Chaoyang District, Beijing 100012, China}
\nocollaboration



\begin{abstract}
We present a catalogue of an H{\sc{i}}-selected sample
 of 1129 low surface brightness galaxies (LSBGs) 
searched from the $\alpha$.40--SDSS DR7 survey.
This sample, consisting of various types of galaxies in terms of luminosity and morphology,
has extended the parameter space covered by the existing LSBG samples
 Based on a subsample of 173 LSBGs which are selected from our entire LSBG sample
 to have the 2-horn shapes of the H{\sc{i}} line profiles,
 minor-to-major axial ratios (b/a) less than 0.6 and signal-to-noise ratio (S/N) of H{\sc{i}}
 detection greater than 6.5, we investigated the Tully-Fisher relation (TFr) of LSBGs
 in the optical $B$, $g$ and $r$ bands and near-infrared $J$, $H$ and $K$ bands as well.
 In optical bands, the LSBG subsample follows the fundamental TFr
which was previously defined for normal spiral galaxies. 
In NIR bands, the TFrs for our LSBG subsample are slightly different from 
the TFrs for the normal bright galaxies. This might be due to
the internal extinction issue. 
 Furthermore, the mass-to-light ratio ($M/L$), 
 disk scale length ($h$) and mass surface density ($\bar{\sigma}$)
 for our LSBG subsample were deduced from the optical TFr results. 
 Compared with High Surface Brightness Galaxies(HSBGs), our LSBGs 
 have higher $M/L$, larger $h$ and lower $\bar{\sigma}$ than HSBGs.
 \end{abstract}



\section{Introduction} \label{sec:intro}
Low Surface Brightness Galaxies (LSBGs) are galaxies that have central surface brightness fainter than the night sky ($\sim$ 22.5 $B$ mag arcsec$^{-2}$ )\citep[e.g.][]{Impey1997,Impey2001,Ceccarelli2012}. Comparing with their High Surface Brightness Galaxy (HSBG) counterparts, they are gas-rich, metal-poor ($\leq 1/3$ solar abundance; \citet{McGaugh1994}), dust-poor \citep[e.g.,][]{Matthews2001}, and with diffuse, low-density stellar disks,  \citep[e.g.][]{de Blok1996,Burkholder2001,O'Neil2004,Trachternach2006}, which indicates that they are low in star formation and insufficiently evolved \citep[e.g.,][]{Das2009,Galaz2011,Lei2018}. 
Usually, dwarf galaxies can mostly satisfy the definition of LSBGs, so dwarf galaxies make a great contribution to the LSBG family. Meanwhile, some of the giant galaxies have very extended and thus faint disk. LSBGs between dwarf and giant types are more similar to late-type spiral galaxies in general properties \citep{Chung2002}. These main types of LSBGs are proposed to have fundamentally different formation and evolutionary histories from each other.  

LSBGs contribute 30\%--60\% to the number density of local galaxies \citep{McGaugh1995,McGaugh1996,Bothun1997,O'Neil2000,Trachternach2006,Haberzettl2007} and $\sim$ 20\% to the dynamical mass of galaxies \citep{Minchin2004}, making them cosmologically significant. However, due to the observational capability and selection effects, LSBGs are seriously underrepresented in the present-day galaxy catalogues, which inevitably leads a bias (towards HSB galaxies) to our understanding of the galaxy formation and evolution. More LSBG samples are necessary. 
The first LSBG samples were mainly composed of LSBGs in the bright end of surface brightness \citep{McGaugh1994,de Blok1995,Schombert1992,Impey1996}. Afterwards, \citet{Zhong2008} derived a large sample of 12282 LSBGs from the main galaxy sample of SDSS DR4 \citep{Adelman-McCarthy2006}, with dwarf galaxies removed. \citet{Williams2016} discovered LSBGs which have fainter surface brightness ($\mu_{0}$(B) $\sim$ 24 -- 28 mag arcsec$^{-2}$) by using SDSS $gri$ combined images within the GAMA equatorial fields \citep{Driver2011,Liske2015}. As LSBGs are mostly rich in H{\sc{i}} gas, the extragalactic H{\sc{i}} surveys, such as the Arecibo Legacy Fast Arecibo L-band Feed Array (ALFALFA; \citet{Giovanelli2005}), are good databases for searching LSBGs. So we have established a sample of 1129 LSBGs from the combination of 40$\%$ data release of ALFALFA ($\alpha$.40; \citet{Haynes2011}) and SDSS DR7 \citep{Abazajian2009} imaging survey. The sample establishment, including the key steps of sky subtraction and surface photometry, have been detailedly reported in \citet{Du2015}, which is the first paper of a series of our papers on LSBG studies. In the current paper, we expect to release the catalogue of the $\alpha.$40-SDSS DR7 LSBG sample.

Additionally, Tully-Fisher relation (TFr), an empirically linear correlation between the rotational velocity and the total absolute magnitude, was first discovered for HSB spiral galaxies in \citet{Tully1977}. The TFr has been widely studied for various samples of HSB spiral galaxies in optical bands \citep{Tully1977,Bottinelli1986,Fouque1990,Sprayberry1995,Courteau1997,Pizagno2007}. Later, 
because near-infrared (NIR) luminosity suffers less from Galactic and internal extinction than optical luminosity, 
and is more sensitive to stellar mass and thus correlates more tightly with the rotational velocity, 
many TFr studies have been performed in NIR passbands \citep{Burstein1982,Freedman1990,Liu2008,Theureau2007,Masters2008,Tiley2016}, 
in mid-infrared passbands \citep{Sorce2012,Sorce2013}, and even in multi-wavelength passbands from optical to NIR \citep{Sakai2000, Bell2001,Ponomareva2017}. 
Recently, the baryonic TFr, a relation between the rotational velocity and the total baryonic mass (including both stellar and gas mass), has received a lot of attention \citep{McGaugh2000,McGaugh2005,Begum2008,Trachternach2009,Gurovich2010,McGaugh2012,Zaritsky2014,Papastergis2016}. 
So far, TFr studies for LSBGs are deficient due to the lack of LSBG samples \citep{Zwaan1995,Sprayberry1995,O'Neil2000,Chung2002}. 
The TFr studies are all performed in optical passbands. \citet{Zwaan1995} gathered 42 LSBGs from literature, and showed that their LSBGs followed 
the fundamental TFr defined for HSB spiral galaxies. This draws a picture that LSBGs may not be fundamentally different from normal HSBGs. 
However, basing on a homogeneous sample of 43 LSBGs, \citet{O'Neil2000} claimed that their LSBGs did not show any significant correlation 
between the velocity width and the optical absolute magnitude. This conclusion was later proved by \citet{Chung2002}. In this paper, we expect to investigate the TFr of LSBGs based on our own LSBG sample. 

In this paper, we briefly introduce the establishment of our LSBG sample in \S 2.1 
and release the complete catalogue in \S 2.2. We perform a luminosity classification in \S 3.1
and a morphological classification for our LSBG sample by machine learning in \S 3.2. 
The optical and radio properties are studied in \S 4. 
In \S 5, we study the optical and near-infrared (NIR) TFr for a subsample from our entire LSBG sample. 
The strength and weakness of this LSBG sample are discussed in \S 6.1.
The mass-to-light ratios, disk scale length and mass surface density deduced
from the TFr results on the basis of our LSBG subsample
are discussed in \S 6.2$\sim$6.4.
Finally, we summarize the paper in \S 7.  
Through this paper, 
the distances that we used to convert angular sizes to physical (kpc) sizes 
are directly the ones listed in the $\alpha$.40 catalogue \citep{Haynes2011}, 
which adopted the Hubble constant to be H$_{0}$=70 km~s$^{-1}$~Mpc$^{-1}$. 
Magnitudes in this paper are all in the AB magnitude system. 

\section{Our $\alpha$.40--SDSS DR7 LSBG sample} \label{sec:cata}
\subsection{A brief review of sample establishment}
In \citet{Du2015}, we have constructed a sample of LSBGs 
which is searched from the $\alpha.$40-SDSS DR7 survey combination. 
We just make a brief introduction of the sample selection in this section
 because details have already been reported in \citet{Du2015}.

First of all, we reconstructed the sky background maps for 
SDSS DR7 $g$- and $r$-band images for each galaxy of
the entire $\alpha.$40-SDSS DR7 sample (12,423 galaxies).
The method we used for reconstructing sky background map
is to fit the sky pixels on the object-subtracted image 
row-by-row and column-by-column \citep{Wu2002,Zheng1999},
which is a better method of constructing sky background for 
bright galaxies with faint extended outskirts
and in particular for LSBGs \citep{Du2015}.
After subtracting the sky background of galaxies,
we made surface photometry by using SExtractor \citep{Bertin1996} 
and made radial profile fitting by using Galfit \citep{Peng2002}.
Afterwards, on the basis of the photometric and fitting results,
the $g-$ and $r-$band central surface brightnesses ($\mu_{0}$) were calculated
and then converted to the $B-$band values ($\mu_{0}$(B)). 
Finally, 1129 galaxies which have $\mu_{0}(B)$ fainter than 22.5 mag arcsec$^{-2}$ 
and minor-to-major axial ratios (b/a) 
in both $g$ and $r$ bands larger than 0.3 
were selected to be our LSBG sample. It is an H{\sc{i}}-selected LSBG sample,
of which the data reduction, sky subtraction, surface photometry, profile fitting,
selection criteria and environment properties have been detailedly reported and
studied in \citet{Du2015}. In this paper, we aim to make the catalogue
of this LSBG sample released to the public who might be interested in this sample
and also investigate the Tully-Fisher relations 
of LSBGs in the optical and NIR bands based on this LSBG sample.

\subsection{Catalogue}
We present the entire catalogue of our LSBG sample in this paper.
A short list of the catalogue is shown in Table ~\ref{tab:cata}
and the long rest list is shown in Table ~\ref{tab:cata1} in Appendix.
The column descriptions of both tables are shown below:

Column 1: galaxy name as it appears in the $\alpha.$40 catalogue.

Column 2: velocity width of the H{\sc{i}} line profile measured at the 50$\%$ level of the peak (stemed from the $\alpha$.40 catalogue).

Columns 3 $\&$ 4:  $g$- and $r$-band magnitudes and errors measured on the sky-subtracted images of galaxies within the automatic aperture (AUTO) by SExtractor. 
As mentioned before, the sky subtraction was performed by using a column-by-column and row-by-row fitting method which was more precise (details in Section 3.1 in \citet{Du2015}).
These magnitudes have already been corrected for the Galactic extinction 
by using the dust maps of \citet{Schlegel1998}(SFD98). 
It is worth noting that SExtractor has computed several types of magnitudes:
isophotal, corrected isophotal, fixed-aperture, automatic aperture and petrosian magnitudes.
Thereinto, the automatic aperture, inspired by Kron's ``first moment'' algorithm (see details in \citet{Kron1980}),
is an flexible and accurate elliptical aperture whose elongation $\epsilon$ and position angle $\theta$ are defined by the second order moments of the object's light distribution. Then, within this aperture, the characteristic radius $r_{1}$ is defined as that weighted by the light distribution function ($r_{1}$= $\frac{\sum rI(r)}{\sum I(r)}$).  \citet{Kron1980} and \citet{Infante1987} have verified that for stars and galaxy profiles convolved with Gaussian seeing, more than 90$\%$ of the flux is expected to lie within a circular aperture of radius $kr_{1}$ if k=2, almost independently of their magnitudes. This picture remains unchanged if they consider an ellipse with $\epsilon kr_{1}$ and $\frac{kr_{1}}{ \epsilon}$ as the principal axes. By choosing a larger k=2.5, more than 96$\%$ of the flux is captured within the elliptical aperture. So, the AUTO magnitudes are intended to give the most precise estimate of ``total magnitudes", at least for galaxies.  
More details about the Kron radius and the automatic aperture photometry used in SExtractor have been reported in SExtractor manual and \citet{Kron1980,Infante1987,Bertin1996}.
For our automatic photometry using SExtractor, we keep k=2.5 that is also the default setting of SExtractor \citep{Bertin1996} and our following studies and analysis would all base on the AUTO magnitudes in this table.

Column 5: $B$-band AUTO magnitude converted from the combination of AUTO magnitudes in both $g$- and $r$ bands by using the transformation formula in \citet{Smith2002} (also listed as Equation (1) below). The errors on $B$-band magnitudes in this table were propagated from the errors on magnitudes in $g$ and $r$ bands by the error propagation method.
\begin{equation}
B = g+0.47(g- r)+0.17
\end{equation}

Columns 6 $\&$ 7:$g$-band major-to-minor axis ratio (b/a) and disk scale length (h(g)) 
which were derived from fitting the sky-subtracted image of the galaxy with a single exponential profile 
by using Galfit \citep{Peng2002}. The details about our fitting processes have been reported in Section 3.3 in \citet{Du2015}). No uncertainties on b/a or h(g) are listed in this table because Galfit did not provide any errors for these outputs. Galfit gave the disk scale length in angular size and we converted the angular size to the physical size (kpc) by using the distance information from the $\alpha$.40 catalogue \citep{Haynes2011}. 

Column 8: $g$-band effective radius (R$_{eff}$(g)) given by SExtractor. SExtractor gave the effective radius in angular size and we then converted the angular size to the physical size by using the distance information from the $\alpha$.40 catalogue. No uncertainties on R$_{eff}$(g) are listed in this table because SExtractor did not provide the uncertainties for this parameter.

Column 9: $g$-band central surface brightness in mag~arcsec$^{-2}$, $\mu_{0,g}$,
 which is calculated from the combination of $g$-band automatic aperture magnitude,$g$, (Column 3), major-to-minor axis ratio,$b/a$(Column 6) and disk scale length,$h(g)$ (Column 7),
 according to the calculation formula (1b) in \citet{Du2015}. 
 As we have no available uncertainties on $b/a$ and $h(g)$, unfortunately we could not provide the uncertainties on $\mu_(0,g)$. 
 Additionally, the $g$-band effective radius and central surface brightness in columns 8 $\&$ 9 are listed in this table only because they might be useful for readers who would like to pick out candidates of Ultra Diffuse Galaxies(UDGs) from this LSBG sample. 

Column 10: $B$-band central surface brightness in mag~arcsec$^{-2}$, $\mu_{0,B}$.
This value was converted from $\mu_{0,g}$ (Column 9) and $\mu_{0,r}$ according to the transformation formula in \citet{Smith2002} which was also listed as formula (1c) in \citet{Du2015} for convenience. 
$\mu_{0,r}$ was calculated using the same method as $\mu_{0,g}$ in Column 9. As we have no available uncertainties on both $\mu_{0,g}$ and $\mu_{0,r}$, we can not provide any uncertainties on $\mu_{0,B}$.

Columns 11$\sim$ 14: Besides the central surface brightness based on the AUTO magnitudes (Column 10), we also list the $B$-band central surface brightnesses based on the isophotal (ISO), corrected isophotal (ISOCOR), petrosian (PETRO) magnitudes 
 and the average of the four  (AVE) for each galaxy.
 As we have difficulty in providing the errors on these central surface brightnesses of each type,
 we wish that readers could make their own impression and constraints on
 the uncertainties of the central surface brightness by comparing $\mu_{0}$
 based on different magnitude types.

Column 15: morphological type by decision trees with Random Forest of Machine Learning (see details in \S ~\ref{subsec:class2})

Column 16: luminosity type in terms of $B$-band absolute magnitude. D, G and M denote dwarf LSBGs, giant LSBGs and moderate-luminosity LSBGs (see details in \S ~\ref{subsec:class1}).

More useful parameters, such as ra, dec, distance, radial velocity and the H{\sc{i}} mass, 
can be easily accessed from the $\alpha$.40 catalogue \citep{Haynes2011} or the entire ALFALFA catalogue \citep{Haynes2018}.

\begin{longrotatetable}
\begin{deluxetable*}{llllllllllllllll}
\tablecaption{A subset of the catalogue of our LSBG sample from the $\alpha$.40--SDSS DR7.\label{tab:cata}}
\tablewidth{750pt}
\tabletypesize{\scriptsize}
\tablehead{
\colhead{AGC} &\colhead{W50}  &
\colhead{g} & \colhead{r} &\colhead{B} & \colhead{b/a} &\colhead{h(g)} &\colhead{R$_{eff}$(g)} &
\colhead{$\mu_{0,g}$} & \colhead{$\mu_{0,B}$} & \colhead{$\mu_{0,B}$} & \colhead{$\mu_{0,B}$} & \colhead{$\mu_{0,B}$} & 
\colhead{$\mu_{0,B}$} & 
 \colhead{T1} &\colhead{T2} \\ 
\colhead{ } &\colhead{(km s$^{-1}$)} &
\colhead{(mag)} &\colhead{(mag)} &\colhead{(mag)} &\colhead{ } &\colhead{(kpc)} &\colhead{(kpc)} &
\colhead{(AUTO)} &\colhead{( AUTO)} &\colhead{ISO} &\colhead{ISOCOR} & \colhead{PETRO} & \colhead{AVE }&
\colhead{ } & \colhead{ }
} 
\startdata
17 &    102$\pm$      2 &14.66$\pm$0.005 &14.20$\pm$0.005 &15.05$\pm$0.008 &0.62 &1.52 &1.87 &23.08 &23.44 &24.96 &23.59 &23.40 &23.85 &S &D
 \\ 
 273 &    157$\pm$     11 &16.38$\pm$0.010 &16.01$\pm$0.010 &16.73$\pm$0.015 &0.53 &3.23 &4.31 &22.26 &22.65 &23.58 &23.04 &22.67 &22.98 &L &M
 \\
 337 &    144$\pm$     13 &15.66$\pm$0.006 &15.36$\pm$0.007 &15.98$\pm$0.010 &0.78 &3.32 &5.67 &22.13 &22.55 &23.63 &23.16 &22.56 &22.98 &L &M
 \\
 1084 &     59$\pm$      3 &16.41$\pm$0.011 &16.03$\pm$0.011 &16.75$\pm$0.017 &0.74 &2.49 &3.20 &23.26 &23.63 &25.31 &24.23 &23.56 &24.18 &L &D
 \\
1211 &    156$\pm$      5 &14.92$\pm$0.005 &14.55$\pm$0.006 &15.27$\pm$0.008 &0.70 &2.68 &3.67 &22.67 &23.14 &24.51 &23.54 &23.10 &23.57 &S &M
 \\
 1362 &    151$\pm$      8 &16.46$\pm$0.012 &16.00$\pm$0.012 &16.85$\pm$0.019 &0.76 &3.99 &5.72 &22.45 &22.90 &23.99 &22.88 &22.88 &23.16 &S &M
 \\
1693 &     95$\pm$      2 &15.33$\pm$0.007 &15.01$\pm$0.008 &15.65$\pm$0.012 &0.84 &3.60 &5.01 &22.86 &23.23 &24.99 &23.18 &23.22 &23.66 &S &M
 \\
2144 &    130$\pm$      7 &15.27$\pm$0.007 &14.79$\pm$0.006 &15.67$\pm$0.010 &0.86 &3.64 &4.99 &22.34 &22.77 &23.40 &22.28 &22.73 &22.79 &S &M
 \\
4528 &     88$\pm$      6 &15.65$\pm$0.006 &15.35$\pm$0.008 &15.96$\pm$0.010 &0.92 &2.92 &5.10 &22.39 &22.90 &24.18 &23.59 &22.86 &23.38 &L &M
 \\
4542 &    143$\pm$      2 &14.97$\pm$0.005 &14.52$\pm$0.005 &15.35$\pm$0.008 &0.84 &4.78 &6.73 &22.29 &22.75 &23.43 &22.63 &22.70 &22.88 &S &G
 \\
4626 &    103$\pm$      2 &15.78$\pm$0.009 &15.36$\pm$0.008 &16.15$\pm$0.014 &0.79 &3.25 &3.75 &23.55 &23.67 &25.29 &24.04 &23.60 &24.15 &L &D
 \\
4797 &     73$\pm$      2 &14.13$\pm$0.003 &13.69$\pm$0.004 &14.51$\pm$0.006 &0.77 &2.93 &3.32 &23.13 &23.48 &24.64 &23.51 &23.44 &23.77 &S &M
 \\
5284 &    185$\pm$      2 &15.67$\pm$0.007 &15.24$\pm$0.007 &16.04$\pm$0.011 &0.66 &3.65 &5.17 &22.16 &22.65 &23.63 &22.98 &22.64 &22.97 &S &M
 \\
5633 &    167$\pm$      2 &13.85$\pm$0.004 &13.43$\pm$0.004 &14.22$\pm$0.006 &0.54 &2.91 &3.52 &22.29 &22.67 &23.73 &22.89 &22.71 &23.00 &S &M
 \\
5716 &    122$\pm$      2 &15.24$\pm$0.005 &15.06$\pm$0.007 &15.50$\pm$0.008 &0.79 &1.35 &1.90 &22.57 &22.93 &24.10 &23.47 &22.81 &23.33 &L &D
 \\
5758 &    121$\pm$      4 &16.41$\pm$0.010 &16.12$\pm$0.013 &16.71$\pm$0.016 &0.54 &2.16 &2.48 &22.67 &22.99 &24.09 &23.36 &22.99 &23.36 &L &D
 \\
5781 &    166$\pm$      3 &20.92$\pm$0.061 &20.61$\pm$0.071 &21.24$\pm$0.095 &0.44 &2.52 &0.43 &27.27 &27.64 &30.99 &26.79 &27.66 &28.27 &L &D
 \\
5889 &     46$\pm$      2 &13.71$\pm$0.002 &13.26$\pm$0.003 &14.08$\pm$0.004 &0.89 &1.06 &1.49 &22.44 &22.83 &23.51 &22.96 &22.82 &23.03 &S &D
 \\
5999 &     68$\pm$      2 &16.24$\pm$0.009 &15.97$\pm$0.010 &16.54$\pm$0.015 &0.67 &4.56 &4.10 &24.09 &24.47 &26.60 &25.72 &24.35 &25.28 &L &M
 \\
6122 &     75$\pm$      5 &15.53$\pm$0.007 &15.15$\pm$0.008 &15.88$\pm$0.011 &0.94 &4.80 &7.73 &22.44 &22.94 &24.08 &23.18 &22.92 &23.28 &S &G
 \\
6130 &    219$\pm$      6 &16.68$\pm$0.011 &16.16$\pm$0.008 &17.10$\pm$0.018 &0.92 &7.68 &4.28 &24.01 &24.51 &25.05 &24.89 &24.23 &24.67 &S &M
 \\
6248 &     26$\pm$      3 &15.72$\pm$0.009 &15.27$\pm$0.008 &16.10$\pm$0.014 &0.84 &1.25 &1.73 &23.37 &23.67 &25.56 &22.96 &23.58 &23.94 &L &D
 \\
6287 &     62$\pm$      5 &16.59$\pm$0.010 &16.29$\pm$0.013 &16.91$\pm$0.016 &0.59 &4.91 &5.76 &23.10 &23.39 &24.98 &24.38 &23.26 &24.00 &L &M
 \\
6486 &    113$\pm$      3 &15.68$\pm$0.007 &15.30$\pm$0.008 &16.03$\pm$0.011 &0.93 &2.71 &4.15 &22.86 &23.15 &24.63 &23.88 &23.13 &23.70 &S &M
 \\
6599 &    118$\pm$      2 &15.18$\pm$0.005 &14.83$\pm$0.006 &15.52$\pm$0.008 &0.41 &2.28 &2.69 &22.48 &22.89 &23.95 &23.41 &22.76 &23.25 &S &D
 \\
6647 &    207$\pm$      2 &15.01$\pm$0.005 &14.53$\pm$0.006 &15.41$\pm$0.008 &0.82 &5.47 &7.20 &22.10 &22.68 &23.40 &22.83 &22.63 &22.89 &S &G
 \\
6669 &     65$\pm$      4 &15.84$\pm$0.007 &15.68$\pm$0.009 &16.08$\pm$0.011 &0.31 &2.26 &1.72 &23.86 &24.05 &26.91 &26.18 &24.05 &25.30 &L &D
 \\
6717 &     61$\pm$      1 &15.16$\pm$0.006 &14.82$\pm$0.006 &15.49$\pm$0.010 &0.79 &4.22 &4.96 &23.37 &23.70 &25.34 &24.35 &23.67 &24.26 &S &M
 \\
6807 &    140$\pm$      2 &15.86$\pm$0.006 &15.61$\pm$0.008 &16.15$\pm$0.010 &0.87 &2.42 &4.04 &22.43 &22.88 &24.21 &23.84 &22.87 &23.45 &L &M
 \\
6980 &     36$\pm$      3 &15.96$\pm$0.007 &15.85$\pm$0.010 &16.18$\pm$0.011 &0.48 &2.98 &3.39 &22.56 &22.86 &24.26 &23.95 &22.80 &23.47 &L &M
 \\
7003 &    127$\pm$      1 &16.22$\pm$0.010 &15.82$\pm$0.011 &16.58$\pm$0.016 &0.46 &1.65 &1.77 &23.19 &23.52 &25.14 &24.13 &23.49 &24.07 &L &D
 \\
7038 &     90$\pm$      1 &15.90$\pm$0.007 &15.56$\pm$0.008 &16.24$\pm$0.011 &0.61 &0.82 &0.86 &23.53 &23.80 &26.11 &24.92 &23.78 &24.65 &L &D
 \\
7138 &    145$\pm$      2 &14.51$\pm$0.004 &14.12$\pm$0.003 &14.87$\pm$0.006 &0.64 &2.75 &3.57 &22.11 &22.56 &23.13 &22.82 &22.54 &22.76 &S &M
 \\
7149 &    131$\pm$      3 &15.10$\pm$0.005 &14.63$\pm$0.005 &15.49$\pm$0.008 &0.71 &1.83 &2.40 &22.19 &22.61 &23.22 &22.46 &22.54 &22.71 &S &D
 \\
7150 &     44$\pm$      1 &15.53$\pm$0.007 &15.35$\pm$0.009 &15.79$\pm$0.011 &0.58 &1.15 &1.28 &22.71 &23.05 &24.12 &23.49 &22.96 &23.41 &L &D
 \\
 \enddata
\end{deluxetable*}
\end{longrotatetable}

\section{Classification of the LSBG sample} \label{sec:class}
\subsection{Luminosity classification}\label{subsec:class1}
The surface brightnesses of galaxies show a continuous distribution, 
so the traditional definition of an LSBG \citep{Impey1997} based on a simple threshold in central surface brightness is in fact unphysical. 
This broad definition means that the existing LSBG catalogues must include LSBGs of different types. 
To date, three main types have been discovered in the LSBG regime. 
They are dwarf LSBGs \citep{Sandage1984, Impey1988}, giant LSBGs \citep{Bothun1987, Bothun1990, Sprayberry1993}, and we
call the LSBGs with luminosities between dwarfs and giants the 'moderate-luminosity' LSBGs. Majority of existing LSBG catalogues
include a large fraction of moderate-luminosity LSBGs, such as \citet{Impey1996,Zhong2008} 

(1) dwarf LSBGs: Generally, galaxies with M$_{B}>$-17~mag are considered as dwarf galaxies (\citet{Dunn2010} and references therein), which can easily satisfy the definition for LSBGs. 
Among the dwarf LSBGs, a population of Ultra Diffuse Galaxy (UDG) \citep{van Dokkum2015}
is special because they have faint ($\mu_{g}$(0)$ >$ 24 mag arcsec$^{-2}$) and very extended disks (effective radius, $R_{eff} >$1.5~kpc).  

(2) giant LSBGs: They are bright (M$_{B}>$-19~mag)\citep{Sabatini2003}, 
but their disk surface brightness are quite faint 
because their disks are very extended. 
For example, the disk of Malin-1, an extreme giant LSB galaxy, has extended to $\sim$ 160~kpc in diameter \citep{Galaz2015,Boissier2016}. Compared with the dwarf LSBGs, giant LSBGs are often redder in color.

(3) morderate-luminosity LSBGs:  
LSBGs with moderate luminosity (-19 $<$ M$_{B} <$~-17~mag) are more similar to late-type spiral galaxies 
in appearance and general properties, except for the overall low surface brightness and surface density. 
Up to now, it is proposed that different types of LSBGs 
may have distinct formation and evolutionary histories 
from each other \citep{Matthews2001}.

\begin{figure}[ht!]
\epsscale{1.3}
\plotone{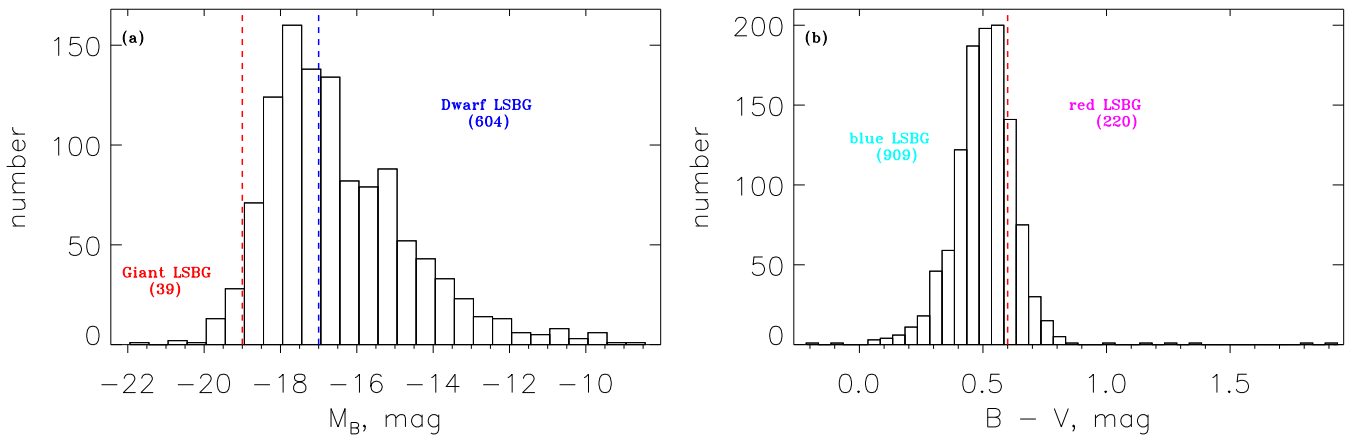}
\caption{Histogram distributions.The $B$-band absolute magnitude (panel (a)) and
and $B$-$V$ color (panel (b)) distributions for the entire LSBG sample are shown. 
In panel (a), the red dashed line represents M$_{B}$=-19.0~mag that distinguishes giants apart from other galaxies, 
and the blue dashed line represents M$_{B}$=-17.0~mag that distinguishes dwarfs from other galaxies. 
Galaxies between the two lines (-19.0$<$M$_{B}<$-17.0) are considered as galaxies with moderate luminosities.
In panel (b), the red dashed line of B-V=0.6~mag is used to divide LSBGs into blue and red.  \label{fig:fig_class}}
\end{figure}

In Figure ~\ref{fig:fig_class}(a), We classified the galaxies in our LSBG sample in terms of $B$-band absolute magnitude. Out of the total 1129 galaxies , 39 galaxies are giants (M$_{B} \leq$-19.0~mag), 
607 galaxies are dwarfs (M$_{B} \geq$-17.0~mag),
among which 28 galaxies are UDG candidates with $\mu_{g}$(0)$ >$ 24 mag arcsec$^{-2}$ and $R_{g,e} >$1.5~kpc.
and the rest 483 galaxies are moderate-luminosity LSBGs (-19.0$<$M$_{B}<$-17.0~mag)  . 
Additionally, in $B$-$V$ color, our sample is dominated by blue galaxies
since it has 909 blue ($B$-$V \leq$ 0.6~mag) and 220 red galaxies ($B-V>$0.6~mag)(Figure ~\ref{fig:fig_class}(b)).

\subsection{Morphological Classification by Machine Learning}\label{subsec:class2}
For a statistical information of the morphologies of our LSBG sample,
we made morphological classification for our sample using Machine Learning Technique. 
Among various machine learning classifiers, we chose the classifier of decision trees with Random Forests 
which has been proved to have the highest accuracy for classification  \citep{Dobrycheva2017}.

Above all, we should derive a training sample to train the classifier.
There is a catalogue of 14,034 SDSS galaxies with detailed visual classifications \citep{Nair2010}, 
covering morphological types from T$=$-5 to 10 (E to Im) in de Vaucouleurs system. 
We randomly selected 80$\%$ of galaxies from this catalogue as the training sample
and the rest 20$\%$ as the test sample.  

First of all, the attributes of redshift (z), $g$-band magnitude ($g$) and absolute magnitude( M$_{g}$), 
$r$-band magnitude($r$), $g$-$r$ color, and the concentration index (R$_{90}$/R$_{50}$) of galaxies in the training sample were adopted to train the Random Forests Classifier model.
The trained Classifier was then fed into the test sample to predict the morphological types and test the accuracy. The accuracy of the trained classifier was 82.5$\%$ for the total test sample and 88.8$\%$ for E type (E, S0), 82.8$\%$ for S type (Sa, Sab, Sb, Sbc, Sc, Scd), 31.4$\%$ for L type (Sd, Sm, Im) and 12.8$\%$ for U type (unknown) of the test sample. 
Secondly, we adopted this trained Random Forests classifier to predict the morphological types (E, S, L, or U) of our LSBG sample. The results from the classifier show that out of the total 1129 galaxies in our sample, 950 galaxies (84.1$\%$) are classified into L type, 151 galaxies (13.4$\%$) are classified into S type, 2 galaxies (0.2$\%$) are classified into E type and 26 galaxies (2.3$\%$) are classified into U type,
which indicates that our LSBG sample are dominating by the late-type galaxies in morphology.
For a general review, the morphological classification results are listed in the column of T$_{M}$ in Table ~\ref{tab:cata}. 

\begin{figure}[ht!]
\epsscale{0.6}
\plotone{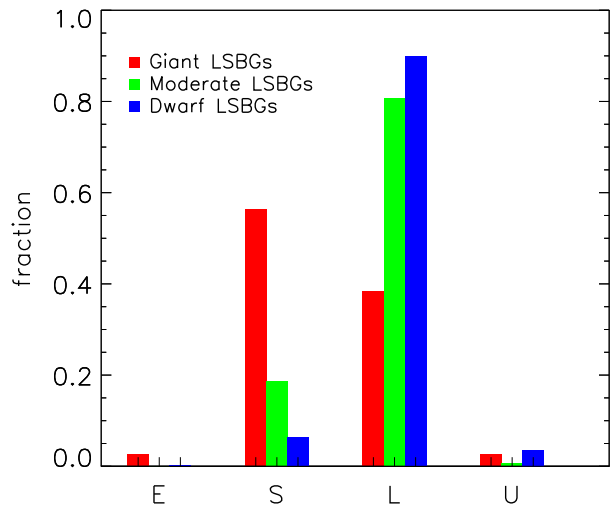}
\caption{Distributions of different morphological types (E, S and L) for giant (red), moderate-luminosity (green) and dwarf (blue) LSBGs of our sample. E represents E and S0 types. S represents Sa, Sab, Sb, Sbc, Sc and Scd types. L represents Sd, Sm and Im types. U represents the unknown morphological type since the morphological type for those galaxies could not be successfully recognized from the Machine Learning technique. \label{fig:lumi_morph}}
\end{figure}

Additionally,  in Figure ~\ref{fig:lumi_morph}, we checked the fraction of galaxies of E, S and L types for our LSBGs of different luminosity types of giant, moderate-luminosity and dwarf LSBGs classified in \S ~\ref{subsec:class1}. For giant LSBGs (red), the morphological types of galaxies are dominated by earlier morphological types (S). For moderate-luminosity LSBGs (green), the morphological types of galaxies are composed of less S and more L types. For dwarf LSBGs (blue), the morphological types of galaxies are absolutely dominated by very late morphological type (L). Therefore, for our LSBG sample of different luminosity types from giant to dwarf, the dominance of morphological types have changed from earlier types (S) to later types (L). So the morphological classifications by Machine Learning technique agrees statistically with the luminosity classification for the three luminosity types of LSBGs in our sample. 

\section{Optical and Radio Properties} \label{sec:property}
\begin{figure}[ht!]
\epsscale{1.3}
\plotone{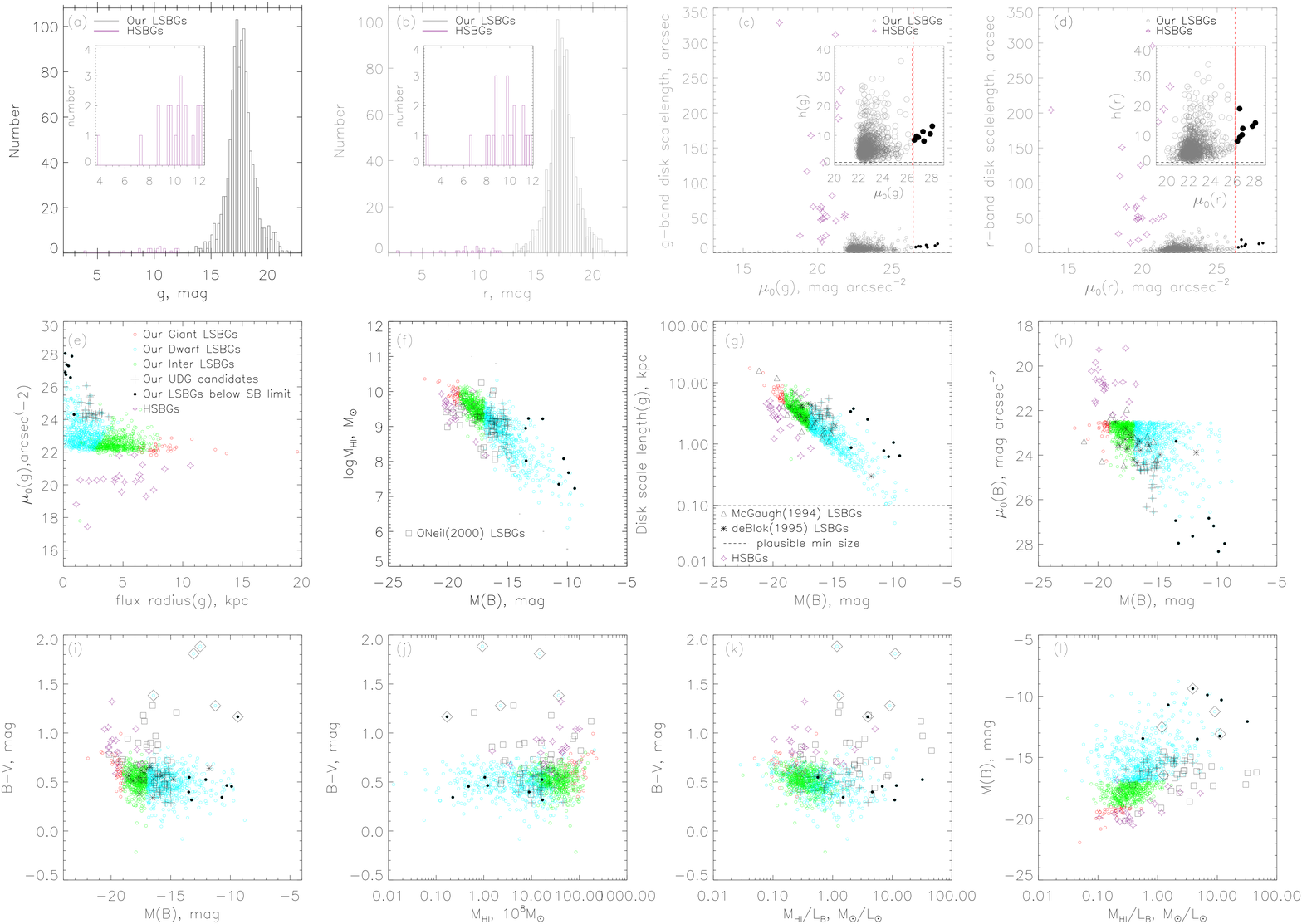}
\caption{Optical and radio properties of our LSBG sample (open circles),compared with HSBGs (purple stars) from \citet{Ponomareva2017} and \citet{Ponomareva2017b}
and other available LSBGs from \citet{O'Neil2000} (open squares),\citet{McGaugh1994}(open triangles) and \citet{de Blok1995}(asterisks). 
In all panels, the cyan, red and green open circles represent dwarf, giant and moderate-luminosity LSBGs in our sample. 
The overlaid black filled circles mark the 7 galaxies in our sample for which we have low confidence in their measurements of central surface brightness.
The over-plotted black plus symbols represent candidates of Ultra Diffuse Galaxies in our LSBG sample.
In panels (a) and (b), the magnitudes which were measured by using SExtractor AUTO aperture in $g$ and $r$ bands for our entire LSBG sample are shown. In panels (c) and (d), the red dashed lines represent the 3$\sigma$ surface brightness limitings of $g\sim$26.4 mag/arcsec$^{2}$ and $r\sim$26.2 mag/arcsec$^{2}$ (within circular apertures of radii of 12$\arcsec$) for the SDSS imaging survey. The black dashed lines represent the half of the maximum PSF width for the SDSS images. Within these 4 panels, the zoom-in plots are shown to show the details of data distribution in small subregions. In panel (e), the flux radius versus the $g$-band central surface brightness of our entire LSBG sample are shown. In panel (f), the $B$-band absolute magnitude versus H{\sc{i}} mass are shown. In panel (g), $B$-band absolute magnitude versus the optical disk scale length are shown and the black dashed line is a widely accepted lower size cutoff for a galaxy. In panel (h), the  $B$-band absolute magnitude versus central surface brightness are shown. In panels (i)$\sim$(l), our LSBGs which have very red color ($B$-$V >$ 1.0 mag) but are low in luminosity (M(B) $>$-17.0 mag) and gas-rich (M$_{H{\sc{i}}}$/L$_{B}>$  1.0) are over-marked by open diamonds.
 \label{fig:fig_para}}
\end{figure}

We show the optical and radio properties of the LSBG sample in Figure ~\ref{fig:fig_para}. 
In each panel, dwarf, moderate-luminosity and giant LSBGs in our sample
are represented by the cyan, green and red circles. 
Additionally, we compare our LSBG sample with the 21 HSBGs (purple stars) 
which have optical band photometry from \citet{Ponomareva2017} and \citet{Ponomareva2017b}
and also with the available LSBGs from \citet{McGaugh1994}(open triangles),
\citet{O'Neil2000} (open squares) and \citet{de Blok1995}(asterisks) in the figure.

We show the $g$- and $r$-band AUTO magnitude distributions for both our LSBG sample (black) and the HSBGs (purple) in panels (a) and (b).
In order to show the distribution of the HSBGs clearly, the zoom-in windows are also plotted within each panel.
The $g$-band magnitude of our LSBGs peaks at 17.57 mag and has an Gaussian FWHM of 2.05 mag in $g$ band.
The $r$-band magnitude of our LSBGs peaks at 17.26 mag and has an Gaussian FWHM of 2.17 mag in $r$ band. 
Comparing with the HSBGs, our LSBGs are entirely much fainter.

In panels (c) and (d), both our LSBG sample and the HSBGs are shown in the size - central surface brightness plane.
Comparing with our LSBGs, the HSBGs are systematically brighter in central surface brightness and larger in apparent size.
As the Point Spread Function (PSF) widths of the SDSS $g$- and $r$-band images are statistically 1.0$\sim$2.0 $\arcsec$,  
galaxies with very small measured disk scale length (below $h=$~1.0$\arcsec$ line) are better to be considered to be spurious. 
In order to show the distribution of LSBGs and the $h=$~1.0$\arcsec$ line clearly, we over-plot the zoom-in windows within panels (c) and (d).
Fortunately, no galaxies in our sample have measured disk scale lengths less than ~1.0$\arcsec$. 
For the SDSS survey, \citet{Kniazev2004} 
found that the 3$\sigma$ limiting surface brightness goes down to $\sim$26.4~mag/arcsec$^{2}$ 
in $g$ band and goes down to $\sim$26.2~mag/arcsec$^{2}$ in $r$ band in a circular aperture of a radius of 12$\arcsec$.
Therefore, we should have lower confidence in our measured $\mu_{0}$ values 
for those galaxies which have measured $\mu_{0,g}$ fainter than 26.4~mag/arcsec$^{2}$ or $\mu_{0,r} $ fainter than 26.2~mag/arcsec$^{2}$ 
but the disk scale lengths are measured to be smaller than 12 $\arcsec$. 
Furthermore, for those galaxies which are extended to have measured disk scale lengths larger than 12 $\arcsec$, 
it is logically reasonable for them to have $\mu_{0,g}$ fainter than 26.4~mag/arcsec$^{2}$ or $\mu_{0,r}$ fainter than 26.2~mag/arcsec$^{2}$.
It mainly depends on the sizes of those galaxies that how faint their central surface brightness could reach? 
From the literature, \citet{Pohlen2006} found that the 3$\sigma$ limiting surface brightness
of the SDSS images could go down to $\sim$27.0 ~mag/arcsec$^{2}$ at a very large circular radius of 150$\arcsec$ in both $g$ and $r$ bands. Therefore, the central surface brightness are not expected to be measured fainter than $\sim$27.0 ~mag/arcsec$^{2}$ for our LSBGs 
which have disk scale length measured to be far less than 150$\arcsec$.
So, for our LSBGs which have disk scale length measured to be larger than 12$\arcsec$ but smaller than 150$\arcsec$, the $\mu_{0}$ measurements for these galaxies that are measured to be fainter than or around $\sim$27.0~mag/arcsec$^{2}$ might be unreliable. 
For a reminder, all of the 7 galaxies with such unreliable measurements of $\mu_{0}$ were marked (as black filled circles) to the right of 
the 3$\sigma$ limiting line (red dashed lines) that was extracted by \citep{Kniazev2004}
for the SDSS imaging survey in Figure ~\ref{fig:fig_para}(c) and (d). 
For emphasis, they were still marked in the following panels.
For further reminders, in Table ~\ref{tab:cata}, the AGC names of these 7 galaxies 
are labeled by adding a star symbol. 
It is necessary to note ahead of time that these 7 galaxies that have unreliable measurements of
central surface brightness would be excluded in the Tully-Fisher relation studies later in \S ~\ref{sec:TF}.

In panel (e),it shows that LSBGs and HSBGs almost cover the similar range in physical size. 
For the LSBG sample itself, it shows that LSBGs which have higher luminosities tends to be larger in physical size. 
Additionally, there are 24 dwarf LSBGs (over-plotted with black plus) which can be candidates of Ultra Diffuse Galaxies (UDGs) 
because they have $\mu_{0}$($g$) $>$ 24.0 mag arcsec$^{-2}$ and effective radius $R_{eff}$(g)$>$1.5 kpc \citep{van Dokkum2015}.
 
In panels (f) and (g), HSBGs (purple stars) are at the bottom edge or below the LSBGs (circles), 
implying that LSBGs usually have larger physical size and have higher H{\sc{i}} mass than the HSBGs with the same level of luminosity.
For the LSBG sample itself, it shows that our LSBGs which have fainter luminosities tends to be 
lower in H{\sc{i}} mass and smaller in size.
This phenomenon for our LSBGs agrees with that for other LSBG samples from \citet{O'Neil2000} (black open squares) and \citet{McGaugh1994} (black open triangles). 
Comparatively, for our sample, giant LSBGs (red circles) have the highest H{\sc{i}} masses and largest sizes, then the moderate-luminosity LSBGs (green open circles), 
and the dwarf LSBGs have the lowest H{\sc{i}} masses and smallest sizes.
It is worth noting that in panel (g), 
an outlier with disk scale length smaller than $\sim$0.1 kpc 
(the black dashed line which is a widely accepted lower size cutoff for a galaxy) 
deviates from most of the LSBGs in size. 
We checked this outlier (AGC~748778) and 
found that it is very nearby (radial velocity=258~km/s), 
has a very low H{\sc{i}} mass (M(HI)=10$^{6.36}$M$_{\odot}$) and a very narrow width of the H{\sc{i}} line profile,
and appears to be extremely faint, irregular and small in SDSS $gri$ combined images.
Such a galaxy image was not expected to be well fitted by an exponential profile model by using Galfit. 
So, the disk scale length measured for this outlier (AGC 748778) might be unreliable. 
However, it will not seriously affect the statistical trend of the sample distribution in the disk scale length -- luminosity plane. 
For a reminder, in Table ~\ref{tab:cata}, the AGC name,748778 which might have unreliable measurements for disk scale length value
 is labeled by adding an asterisk symbol. Additionally, this galaxy would also be excluded in the Tully-Fisher relation fitting later in this paper.

In panel (h), The HSBGs are systematically brighter in both central surface brightness and luminosity than the LSBGs.
By comparison, there are obviously no LSBGs with $\mu_{0}$(B)$< $22.5~mag arcsec$^{-2}$.
This is due to our selection threshold. 
It is interesting that in the $\mu_{0}$(B) -- M($B$) plane, 
there is a void at the lower left corner for our sample and 
another sample of \citet{McGaugh1994} (open triangles). 
This void space should have been populated by very low surface brightness, 
but intrinsically very luminous and large galaxies,
which are in fact the giant LSBGs.
This void might be either due to the low number density of giant LSBGs or 
that our sample actually lacks for LSBGs with bulges due to our selection method
(see more in Section 6.1 or in \citet{Du2015}).

In panels (i)$\sim$ (l), comparing with LSBGs, the HSBGs are systematically redder in color. 
For the LSBGs, while our sample has extended the parameter space covered by the previous samples towards lower luminosity and lower H{\sc{i}} mass, 
the previous samples also cover parameter space that is not covered by our sample, specifically towards redder B-V color and higher H{\sc{i}} mass-to-luminosity ratio. 
The main reason might be that our sample lacks for LSBGs with bulges 
which are generally redder than disk and dwarf galaxies
due to the selection method (see more discussions in Section 6.1 or in \citet{Du2015}),
although five LSBGs (over-plotted with open diamonds) in our sample
appear to be very red ($B$-$V >$ 1.0 mag), low-luminosity (M(B) $>$-17.0 mag) (panel (i))
and gas-rich (M$_{H{\sc{i}}}$/L$_{B}>$  1.0; panels (k) and (l)).
In panel (l),  a clear trend between absolute magnitude and gas-to-star ratio appears to exist for the entire LSBG sample,
showing that fainter LSBGs would be larger in gas-to-star ratio, however, it shows no color--absolute magnitude relation (panel (i)), color--gas content correlation (panel (j)) nor color--gas-to-star relation (panel (k)).

\section{Tully-Fisher relation}\label{sec:TF}
The Tully-Fisher relation (TFr) has been well defined 
between the luminosity and the maximal rotational velocity
for late-type spiral galaxies (with high surface brightness) \citep{Tully1977}.

There are two main methods to measure the maximal rotational velocities, 
V$_{max}$, of spiral galaxies. Firstly, V$_{max}$ could be measured from the
width of the global H{\sc{i}} line profile, as the observed width of the H{\sc{i}} line profile, 
in km/s, gives the observed Doppler broadening due principally to the galaxy's rotation.
In the literature, the width of the H{\sc{i}} line profile at 50$\%$ of peak flux, W$_{50}$, 
has been greatly used as a good measure of the V$_{max}$ of a galaxy to, 
study the TFr of spiral galaxies in the absolute magnitude - W$_{50}$ plane
\citep{Zwaan1995,O'Neil2000,Chung2002}. 
Secondly, V$_{max}$ could be measured from the high-quality rotation curves
which can be derived from spatially-resolved H{\sc{i}} velocity maps of galaxies.
It is clearly demonstrated in \citet{Ponomareva2017} that
the shape of the global H{\sc{i}} line profile may hint at the shape of the rotation curve
of a galaxy in several cases.
Generally, the global H{\sc{i}} line profile can be observationally obtained much faster than
the high-quality H{\sc{i}} velocity map for a galaxy, so the first method which use the line width to measure
the maximal rotation velocity of a spiral galaxy is feasible for large samples of galaxies.
A detailed comparisons in Figure 6 in \citet{Ponomareva2017}
demonstrated that the rotational velocity measured from W$_{50}$ will be underestimated
in comparison with V$_{max}$ derived from the high-quality rotation curve for those dwarf 
galaxies which mostly have slowly rising rotation curves or for those massive spiral galaxies
which mostly have declining rotation curves. However, for those galaxies which have the classical 
double peak (2-horn) profiles, the rotational velocity measured from W$_{50}$ are generally
consistent with V$_{max}$ derived from the rotation curves. Because the 2-horn line profile gives 
an indication that the rotation curve of a galaxy will reach its flat part. In view of this, our entire sample which 
is composed of
a variety of galaxy types from dwarf to giant in luminosity and from disk-like to irregular in morphology,
corresponding to various rotation curves, 
should be carefully refined in terms of H{\sc{i}} line profile for better studying TFr.

  In order to make sure that the W$_{50}$
is a good measure of V$_{max}$ of the galaxy,  
first of all, with the examples of good 2-horn H{\sc{i}} profiles given in
\citet{Courtois2009} as standards, we manually inspected the H{\sc{i}} line profile 
for every LSBG of our entire sample 
and selected those galaxies which have 2-horn H{\sc{i}} line profiles.
For example, we show the 2-horn H{\sc{i}} line profiles of several of our LSBGs to give an idea of our identification for
the 2-horn profiles in Figure ~\ref{fig:HI_profile}. These example LSBGs were chosen to have their signal-to-noise (SNR) of H{\sc{i}} detection ranging from high to low.
Secondly, we are recommended by the $\alpha$.40 release paper \citep{Haynes2011} to have those galaxies with the signal-to-noise ratio (SNR) 
of the H{\sc{i}} line less than 6.5 excluded. Because it is claimed in \citet{Haynes2011} that the detection of H{\sc{i}} with SNR$>$6.5 by ALFALFA is a real detection. 
Thirdly, large uncertainties might be inherent in inclination correction for W$_{50}$ for less inclined galaxies ( nearly face-on, with large minor-to-major axis ratio, b/a). 
This problem could be largely mitigated for largely inclined galaxies (nearly edge-on, with small b/a), in particular for edge-on galaxies, so we decided to further have those less inclined galaxies ($g$-band b/a $gid$ 0.6 ) excluded.
This step is only to mitigate the large uncertainties inherent in inclination correction for W$_{50}$ of less inclined galaxies.

After the above criteria of H{\sc{i}}-line profile shape, the SNR of H{\sc{i}} line and the b/a, 
a subsample of 175 LSBGs were selected from our entire sample. 
This subsample is appropriate for the TFr studies because the key point is that,
for this subsample galaxies,
the W$_{50}$ of the H{\sc{i}} line can accurately measure the maximal rotational velocities.
To make the results more reliable, we removed the galaxies which are members of the 7 galaxies
 that were mentioned in Section 4 to have unreliable measurements of central surface brightness and 
 also removed the galaxy AGC 748778 that was also mentioned in Section 4 to have very small disk scale length
measurements from this subsample. After such removal, our final subsample for TFr studies contains
173 LSBGs. We list these 173 LSBGs in Table ~\ref{tab:lsbg_tf} which includes the AGC name, inclination angle, i, calculated following Equation (3), corrected W$_{50}$ calculated by Equation (2), distance from the released entire ALFALFA catalogue, and the $B$-band absolute magnitude which has been corrected for Galactic and internal extinction following Equation (4), the NIR-band absolute magnitudes which have been corrected only for Galactic extinction. For those galaxies with no UKIDSS observations, we just leave their corrected $J$-, $H$- and $K$-band absolute magnitudes blank in Table ~\ref{tab:lsbg_tf}.

 Although this subsample is only 15$\%$ of the entire sample in size, it includes all of the galaxies in the entire sample that are better suitable for TFr studies because the maximum rotation velocity of these galaxies could be properly measured by the width of the H{\sc{i}} line and these galaxies also suffered less from the uncertainties induced by the inclination correction for the H{\sc{i}} line width.

\begin{longrotatetable}
\begin{deluxetable*}{llllllllllllllll}
\tablecaption{A subset of the catalogue of our LSBG sample from the $\alpha$.40--SDSS DR7.\label{tab:cata}}
\tablewidth{750pt}
\tabletypesize{\scriptsize}
\tablehead{
\colhead{AGC} &\colhead{i} & \colhead{W50$_{cor}$}  &
\colhead{Dist} & \colhead{$B_{cor}$} &\colhead{$J_{cor}$} & \colhead{$H_{cor}$} &\colhead{$K_{cor}$} \\ 
\colhead{ } &\colhead{(deg)} &
\colhead{(km/s)} &\colhead{(Mpc)} &\colhead{(mag)} &\colhead{mag } &\colhead{(mag)} &\colhead{(mag)} 
} 
\startdata
273 &59.4 &182.4$\pm$12.78 & 78.9$\pm$2.30 &-18.84$\pm$0.07 & -- &  --& -- \\
5633 &58.8 &195.3$\pm$ 2.34 & 22.5$\pm$4.20 &-18.68$\pm$0.41 &--  & -- & -- \\
5758 &59.2 &140.9$\pm$ 4.66&  45.4$\pm$4.20 &-17.53$\pm$0.202 &-17.02$\pm$0.227 &-17.41$\pm$0.207 &-17.19$\pm$0.215 \\
6287 &55.4 & 75.3$\pm$ 6.07 & 93.9$\pm$2.40 &-18.88$\pm$0.06 & -- & -- &--  \\
6599 &68.1 &127.2$\pm$ 2.16 & 26.2$\pm$2.30 &-17.78$\pm$0.19 &--  &--  &--  \\
6669 &76.0 & 67.0$\pm$ 4.12&  16.1$\pm$2.10 &-16.05$\pm$0.285 &-15.59$\pm$0.299 &-15.57$\pm$0.295 &-15.41$\pm$0.301 \\
7003 &64.4 &140.8$\pm$ 1.11&  25.5$\pm$4.20 &-16.66$\pm$0.358 &-16.32$\pm$0.373 &-16.46$\pm$0.365 &-16.43$\pm$0.366 \\
7737 &70.9 &109.0$\pm$ 1.06&  16.7$\pm$1.20 &-16.94$\pm$0.158 &-16.15$\pm$0.163 &-16.27$\pm$0.160 &-15.72$\pm$0.165 \\
8030 &67.7 & 57.3$\pm$ 2.16 &  7.8$\pm$2.30 &-14.67$\pm$0.64 & -- & -- &--  \\
8055 &62.8 &100.0$\pm$ 2.25&  16.6$\pm$4.30 &-15.97$\pm$0.563 &-15.74$\pm$0.568 &-15.46$\pm$0.566 &-15.79$\pm$0.567 \\
8276 &62.2 & 76.8$\pm$ 2.26&  16.9$\pm$2.20 &-15.61$\pm$0.283 &-15.28$\pm$0.302 &-15.26$\pm$0.292 &-15.17$\pm$0.296 \\
8762 &59.4 &176.7$\pm$ 3.49&  51.7$\pm$2.20 &-19.27$\pm$0.094 &-18.00$\pm$0.148 &-18.76$\pm$0.100 &-18.43$\pm$0.113 \\
8915 &63.8 &176.1$\pm$ 2.23&  37.3$\pm$2.30 &-17.75$\pm$0.135 &-17.36$\pm$0.151 &-17.46$\pm$0.137 &-17.22$\pm$0.148 \\
9002 &67.9 &103.6$\pm$ 4.32&  60.8$\pm$2.30 &-19.80$\pm$0.084 &-18.90$\pm$0.099 &-19.24$\pm$0.086 &-18.87$\pm$0.089 \\
9063 &76.1 &181.3$\pm$ 3.09&  87.9$\pm$4.20 &-19.98$\pm$0.107 &-18.91$\pm$0.127 &-19.16$\pm$0.111 &-18.71$\pm$0.121 \\
9380 &66.0 &104.0$\pm$ 2.19&  27.8$\pm$4.40 &-17.95$\pm$0.344 &-17.23$\pm$0.351 &-17.33$\pm$0.347 &-16.99$\pm$0.351 \\
10009 &71.7 &130.6$\pm$ 4.21&  32.8$\pm$2.40 &-17.55$\pm$0.161 &-16.58$\pm$0.201 &-17.04$\pm$0.163 &-17.12$\pm$0.166 \\
100350 &61.1 &122.2$\pm$14.84 & 64.2$\pm$4.30 &-17.05$\pm$0.15 & -- & -- & -- \\
101812 &69.7 & 99.1$\pm$ 3.20 & 27.0$\pm$2.30 &-15.06$\pm$0.19 & -- & -- & -- \\
101877 &66.4 &201.9$\pm$ 4.36&  72.9$\pm$2.50 &-18.28$\pm$0.078 &-18.01$\pm$0.169 &-18.18$\pm$0.116 &-18.31$\pm$0.117 \\
101942 &55.2 & 92.6$\pm$ 9.75&  79.7$\pm$2.30 &-17.27$\pm$0.070 &-17.89$\pm$0.129 &-17.45$\pm$0.147 &-16.14$\pm$0.491 \\
102229 &60.6 & 94.1$\pm$ 9.18 & 45.9$\pm$2.20 &-16.34$\pm$0.11 & -- & -- & -- \\
102729 &67.1 & 57.5$\pm$ 6.51 & 65.4$\pm$2.10 &-16.24$\pm$0.08 & -- & -- &--  \\
102900 &61.2 &151.8$\pm$ 4.57 &168.6$\pm$2.20 &-17.42$\pm$0.05 &--  & -- &--  \\
110398 &62.2 &198.9$\pm$ 7.91&  92.3$\pm$2.30 &-19.55$\pm$0.057 &-19.35$\pm$0.090 &-19.33$\pm$0.069 &-18.94$\pm$0.090 \\
113200 &56.9 &120.5$\pm$47.74& 102.3$\pm$2.20 &-17.78$\pm$0.057 &-17.51$\pm$0.320 &-17.62$\pm$0.200 &-17.01$\pm$0.352 \\
113752 &57.0 &108.6$\pm$ 8.35 &173.3$\pm$2.30 &-17.89$\pm$0.05 & -- & -- &--  \\
113825 &73.6 &111.5$\pm$ 9.38 & 52.4$\pm$2.40 &-16.34$\pm$0.11 & -- & -- & -- \\
121174 &63.3 & 88.4$\pm$ 3.36 &  9.7$\pm$2.30 &-12.76$\pm$0.52 & -- & -- & -- \\
122218 &72.1 &103.0$\pm$ 2.10 & 35.9$\pm$2.30 &-15.94$\pm$0.14 & -- & -- & -- \\
122341 &60.8 &195.9$\pm$ 4.58& 156.8$\pm$2.30 &-19.47$\pm$0.041 &-18.74$\pm$0.130 &-19.43$\pm$0.053 &-19.14$\pm$0.068 \\
122874 &62.0 &152.8$\pm$11.32 & 87.8$\pm$2.30 &-17.76$\pm$0.06 & -- &--  & -- \\
122877 &56.2 & 97.5$\pm$ 6.02 & 85.0$\pm$2.30 &-18.08$\pm$0.06 &--  &--  &--  \\
123047 &60.1 & 70.4$\pm$ 9.23 &142.7$\pm$2.40 &-17.97$\pm$0.06 &--  &--  &--  \\
123172 &69.0 &171.4$\pm$ 3.21 & 74.5$\pm$2.30 &-18.43$\pm$0.07 &--  &--  &--  \\
171585 &61.0 &144.0$\pm$ 9.15&  69.1$\pm$2.50 &-17.86$\pm$0.081 &-17.60$\pm$0.105 &-17.76$\pm$0.097 &-17.71$\pm$0.103 \\
174601 &65.7 &189.9$\pm$ 3.29 & 72.7$\pm$2.10 &-19.09$\pm$0.07 &--  &--  &--  \\
181471 &73.5 & 94.9$\pm$ 5.21 & 30.3$\pm$2.40 &-16.44$\pm$0.17 & -- & -- &--  \\
181571 &70.0 & 73.4$\pm$ 3.19&  62.3$\pm$2.30 &-18.08$\pm$0.084 &-17.27$\pm$0.155 &-17.56$\pm$0.106 &-17.11$\pm$0.148 \\
181762 &61.8 &240.6$\pm$24.97& 192.3$\pm$2.40 &-20.60$\pm$0.033 &-20.44$\pm$0.074 &-20.77$\pm$0.043 &-20.34$\pm$0.054 \\
182466 &58.6 & 93.8$\pm$ 2.34 & 29.6$\pm$2.30 &-14.73$\pm$0.17 &--  & -- & -- \\
182467 &59.8 & 39.3$\pm$ 9.25 & 59.6$\pm$2.10 &-15.19$\pm$0.09 &--  & -- & -- \\
182477 &55.3 &102.2$\pm$ 3.65 & 62.2$\pm$2.20 &-16.05$\pm$0.09 & -- & -- &--  \\
182608 &69.6 &116.3$\pm$ 4.27&  67.9$\pm$2.30 &-17.44$\pm$0.080 &-16.86$\pm$0.164 &-17.32$\pm$0.128 &-16.69$\pm$0.232 \\
182924 &64.5 &247.0$\pm$ 7.75& 167.7$\pm$2.60 &-19.05$\pm$0.044 &-19.14$\pm$0.165 &-19.13$\pm$0.187 &-19.57$\pm$0.123 \\
183068 &66.6 &179.8$\pm$ 7.63& 183.4$\pm$2.40 &-20.11$\pm$0.036 &-19.86$\pm$0.091 &-20.39$\pm$0.071 &-19.96$\pm$0.108 \\
183138 &57.4 &151.9$\pm$13.05& 171.2$\pm$2.20 &-20.03$\pm$0.033 &-19.89$\pm$0.070 &-20.18$\pm$0.069 &-20.09$\pm$0.082 \\
183691 &61.5 &172.9$\pm$13.65& 124.8$\pm$2.30 &-18.89$\pm$0.046 &-18.62$\pm$0.122 &-19.02$\pm$0.100 &-18.91$\pm$0.102 \\
184304 &70.1 &175.4$\pm$ 7.44&  78.1$\pm$2.30 &-17.56$\pm$0.072 &-16.97$\pm$0.166 &-17.50$\pm$0.123 &-16.86$\pm$0.182 \\
191817 &65.6 &136.2$\pm$ 2.20&  40.6$\pm$2.30 &-17.07$\pm$0.125 &-16.77$\pm$0.145 &-16.59$\pm$0.141 &-16.36$\pm$0.154 \\
191906 &71.0 &125.8$\pm$ 3.17&  48.6$\pm$2.10 &-16.60$\pm$0.099 &-15.94$\pm$0.262 &-16.19$\pm$0.176 &-15.90$\pm$0.181 \\
191989 &56.6 &130.6$\pm$ 5.99 & 48.9$\pm$2.20 &-17.64$\pm$0.10 & -- & -- & -- \\
192118 &61.5 &182.0$\pm$ 3.41&  83.1$\pm$2.30 &-18.16$\pm$0.064 &-17.75$\pm$0.131 &-17.74$\pm$0.093 &-17.41$\pm$0.113 \\
192165 &57.9 &131.0$\pm$ 7.08& 123.2$\pm$2.30 &-19.23$\pm$0.045 &-19.35$\pm$0.067 &-19.50$\pm$0.070 &-19.31$\pm$0.067 \\
193781 &58.9 &177.4$\pm$ 4.67 &121.5$\pm$2.30 &-18.14$\pm$0.05 & -- & -- & -- \\
193802 &70.4 & 45.6$\pm$ 3.18&  23.5$\pm$2.30 &-13.55$\pm$0.216 &-14.28$\pm$0.303 &-13.98$\pm$0.291 &-13.78$\pm$0.322 \\
193862 &68.4 &180.7$\pm$ 8.60 & 56.6$\pm$2.20 &-17.19$\pm$0.09 & -- &--  &--  \\
193905 &62.5 & 92.4$\pm$ 7.89 &115.6$\pm$2.40 &-17.82$\pm$0.06 & -- &--  &--  \\
194002 &63.7 & 85.9$\pm$ 3.35 & 38.1$\pm$2.20 &-16.34$\pm$0.13 & -- &--  &--  \\
198454 &68.5 & 48.4$\pm$ 6.45 & 22.0$\pm$2.20 &-13.40$\pm$0.22 &--  & -- & -- \\
198456 &69.9 & 60.7$\pm$ 6.39&  29.5$\pm$2.20 &-14.58$\pm$0.167 &-13.79$\pm$0.295 &-14.16$\pm$0.195 &-14.01$\pm$0.228 \\
202040 &64.0 &106.8$\pm$ 4.45&  17.5$\pm$3.30 &-14.63$\pm$0.410 &-14.58$\pm$0.437 &-14.77$\pm$0.420 &-14.12$\pm$0.449 \\
202257 &68.7 & 54.7$\pm$ 2.15&  10.4$\pm$4.30 &-14.77$\pm$0.898 &-14.43$\pm$0.904 &-14.62$\pm$0.902 &-14.60$\pm$0.902 \\
202704 &61.0 &101.7$\pm$ 8.00&  99.2$\pm$2.40 &-18.49$\pm$0.057 &-17.96$\pm$0.176 &-18.30$\pm$0.086 &-18.02$\pm$0.109 \\
203515 &54.9 &227.4$\pm$ 7.33 &100.7$\pm$2.20 &-18.44$\pm$0.05 & -- &--  &--  \\
203667 &56.7 & 86.1$\pm$ 8.37&  56.6$\pm$2.30 &-16.98$\pm$0.091 &-16.89$\pm$0.288 &-16.76$\pm$0.155 &-16.01$\pm$0.312 \\
204112 &71.5 & 95.9$\pm$23.19& 185.7$\pm$2.20 &-20.10$\pm$0.037 &-19.62$\pm$0.082 &-19.79$\pm$0.056 &-19.84$\pm$0.064 \\
205062 &67.1 & 58.6$\pm$ 6.51 & 99.2$\pm$2.40 &-16.16$\pm$0.08 & -- &--  &--  \\
205205 &69.3 &110.1$\pm$ 6.41&  43.7$\pm$2.30 &-15.91$\pm$0.121 &-15.37$\pm$0.245 &-15.52$\pm$0.154 &-14.98$\pm$0.271 \\
205216 &55.9 & 48.3$\pm$ 3.62&  47.5$\pm$2.30 &-15.31$\pm$0.111 &-15.51$\pm$0.243 &-14.85$\pm$0.293 &-13.80$\pm$0.816 \\
205336 &56.5 & 92.3$\pm$14.39& 158.7$\pm$2.40 &-15.28$\pm$0.123 &-16.86$\pm$0.379 &-16.40$\pm$0.447 &-16.60$\pm$0.362 \\
206787 &60.6 &308.9$\pm$ 5.74 &163.9$\pm$2.20 &-20.58$\pm$0.03 & -- &--  & -- \\
208385 &59.6 &131.1$\pm$ 2.32 & 55.1$\pm$2.40 &-17.19$\pm$0.10 & -- & -- & -- \\
214098 &59.9 &164.2$\pm$ 3.47&  96.8$\pm$2.30 &-18.60$\pm$0.056 &-18.21$\pm$0.141 &-18.33$\pm$0.140 &-18.79$\pm$0.101 \\
215197 &62.1 &113.2$\pm$ 7.92&  50.5$\pm$2.10 &-16.68$\pm$0.094 &-16.47$\pm$0.174 &-16.46$\pm$0.134 &-16.10$\pm$0.169 \\
215213 &76.1 & 70.0$\pm$ 1.03&   9.0$\pm$2.30 &-13.61$\pm$0.556 &-13.25$\pm$0.567 &-13.02$\pm$0.564 &-13.03$\pm$0.564 \\
215222 &59.1 &179.4$\pm$ 9.32&  97.3$\pm$2.30 &-18.53$\pm$0.056 &-18.18$\pm$0.119 &-18.61$\pm$0.075 &-17.97$\pm$0.111 \\
215256 &71.7 &110.6$\pm$11.58 & 21.0$\pm$2.30 &-16.49$\pm$0.24 & -- & -- & -- \\
215262 &66.0 & 68.9$\pm$ 6.57 & 17.5$\pm$3.60 &-14.29$\pm$0.45 & -- &--  &--  \\
215265 &64.0 &139.0$\pm$ 7.79& 140.3$\pm$2.20 &-18.58$\pm$0.045 &-18.76$\pm$0.119 &-18.60$\pm$0.102 &-18.44$\pm$0.144 \\
215280 &73.4 & 97.1$\pm$ 4.17 & 17.5$\pm$3.70 &-14.05$\pm$0.46 & -- &--  &--  \\
215283 &59.9 &126.0$\pm$ 3.47 & 90.9$\pm$2.30 &-16.48$\pm$0.07 &--  & -- & -- \\
215630 &55.1 &137.8$\pm$ 6.10&  90.5$\pm$2.30 &-17.79$\pm$0.060 &-17.74$\pm$0.123 &-17.17$\pm$0.157 &-17.15$\pm$0.168 \\
215716 &74.8 &103.6$\pm$13.47&  38.3$\pm$2.40 &-16.62$\pm$0.139 &-15.90$\pm$0.189 &-15.91$\pm$0.166 &-15.89$\pm$0.182 \\
219206 &68.5 &148.4$\pm$ 4.30&  97.6$\pm$2.30 &-17.34$\pm$0.067 &-16.07$\pm$0.389 &-16.27$\pm$0.336 &-16.83$\pm$0.228 \\
219232 &54.9 & 58.6$\pm$12.22& 146.8$\pm$2.10 &-16.53$\pm$0.077 &-16.57$\pm$0.423 &-16.59$\pm$0.201 &-15.86$\pm$0.522 \\
220005 &54.8 & 91.7$\pm$ 0.00 &104.0$\pm$2.30 &-17.73$\pm$0.06 & -- & -- & -- \\
220299 &70.8 & 98.5$\pm$ 4.24&  63.4$\pm$2.10 &-17.27$\pm$0.078 &-16.49$\pm$0.362 &-17.41$\pm$0.170 &-17.05$\pm$0.171 \\
220351 &67.3 & 78.1$\pm$ 3.25&  16.7$\pm$1.20 &-15.84$\pm$0.157 &-15.24$\pm$0.169 &-15.11$\pm$0.162 &-14.97$\pm$0.167 \\
220580 &76.1 &118.5$\pm$18.54&  61.8$\pm$2.30 &-18.23$\pm$0.086 &-17.48$\pm$0.133 &-17.75$\pm$0.098 &-17.60$\pm$0.115 \\
220762 &66.4 & 96.1$\pm$ 3.27&  16.6$\pm$4.30 &-15.81$\pm$0.563 &-15.31$\pm$0.570 &-15.34$\pm$0.565 &-15.07$\pm$0.570 \\
220939 &63.4 & 76.0$\pm$ 3.35&  16.5$\pm$1.20 &-15.35$\pm$0.159 &-14.97$\pm$0.186 &-15.10$\pm$0.164 &-14.63$\pm$0.194 \\
221085 &63.3 & 99.6$\pm$ 1.12&  16.6$\pm$2.40 &-15.16$\pm$0.315 &-15.01$\pm$0.341 &-14.74$\pm$0.332 &-14.89$\pm$0.331 \\
223279 &75.6 &124.9$\pm$ 5.16&  89.1$\pm$2.30 &-17.86$\pm$0.066 &-17.64$\pm$0.168 &-17.44$\pm$0.115 &-17.64$\pm$0.127 \\
223295 &71.6 &159.2$\pm$16.87 &106.8$\pm$2.30 &-17.65$\pm$0.06 &--  &--  &--  \\
223392 &63.6 & 92.7$\pm$ 6.70&  24.2$\pm$4.20 &-15.39$\pm$0.378 &-15.70$\pm$0.402 &-15.33$\pm$0.385 &-15.17$\pm$0.389 \\
224075 &67.0 &151.0$\pm$23.89&  97.5$\pm$2.30 &-18.11$\pm$0.059 &-17.94$\pm$0.188 &-17.85$\pm$0.142 &-17.91$\pm$0.144 \\
224094 &55.8 &128.2$\pm$ 6.05&  30.1$\pm$2.30 &-15.97$\pm$0.167 &-16.16$\pm$0.200 &-15.61$\pm$0.210 &-15.75$\pm$0.205 \\
225297 &67.9 &136.0$\pm$ 6.48&  78.9$\pm$2.10 &-17.66$\pm$0.065 &-18.21$\pm$0.140 &-17.68$\pm$0.134 &-17.30$\pm$0.189 \\
225849 &61.0 & 98.3$\pm$ 5.72&  16.3$\pm$2.30 &-14.81$\pm$0.307 &-14.43$\pm$0.324 &-14.03$\pm$0.349 &-13.76$\pm$0.425 \\
225876 &59.3 & 45.4$\pm$10.47&  16.6$\pm$4.30 &-12.41$\pm$0.565 &-11.49$\pm$1.060 &-11.44$\pm$0.783 &-12.03$\pm$0.680 \\
225992 &64.6 &160.5$\pm$12.18&  90.9$\pm$2.20 &-17.08$\pm$0.063 &-16.86$\pm$0.243 &-16.89$\pm$0.125 &-16.56$\pm$0.219 \\
228097 &59.6 & 78.9$\pm$ 8.12&  88.6$\pm$2.20 &-17.70$\pm$0.060 &-17.86$\pm$0.155 &-18.06$\pm$0.099 &-17.55$\pm$0.154 \\
232871 &56.8 &215.2$\pm$ 8.37& 112.7$\pm$2.30 &-17.97$\pm$0.052 &-18.04$\pm$0.173 &-17.61$\pm$0.137 &-17.68$\pm$0.163 \\
232879 &56.2 &178.2$\pm$14.45&  88.6$\pm$2.30 &-18.60$\pm$0.060 &-18.38$\pm$0.136 &-18.59$\pm$0.081 &-18.42$\pm$0.103 \\
233571 &56.6 &131.8$\pm$ 2.40 & 22.9$\pm$2.20 &-16.15$\pm$0.21 & -- &--  &--  \\
233601 &56.1 &120.4$\pm$ 3.61&  22.3$\pm$2.30 &-15.18$\pm$0.225 &-15.13$\pm$0.257 &-15.30$\pm$0.243 &-14.56$\pm$0.281 \\
233615 &66.9 &101.1$\pm$ 7.61&  52.1$\pm$2.20 &-15.79$\pm$0.100 &-15.44$\pm$0.174 &-15.65$\pm$0.179 &-15.65$\pm$0.186 \\
233635 &75.6 &165.2$\pm$ 4.13 &104.9$\pm$2.30 &-18.49$\pm$0.06 &--  &--  & -- \\
233638 &56.5 & 74.4$\pm$ 7.20 &112.3$\pm$2.30 &-18.03$\pm$0.05 &--  & -- & -- \\
233683 &57.1 & 79.8$\pm$ 3.58&  96.8$\pm$2.10 &-18.04$\pm$0.053 &-17.72$\pm$0.108 &-17.86$\pm$0.078 &-17.84$\pm$0.087 \\
233711 &66.3 &126.7$\pm$ 4.37&  83.5$\pm$2.30 &-17.60$\pm$0.066 &-17.04$\pm$0.186 &-17.32$\pm$0.126 &-17.09$\pm$0.143 \\
238689 &63.4 &116.4$\pm$ 8.95&  57.6$\pm$2.30 &-16.86$\pm$0.091 &-16.68$\pm$0.195 &-16.33$\pm$0.214 &-15.84$\pm$0.358 \\
238751 &58.9 &126.1$\pm$10.51&  96.6$\pm$2.20 &-18.13$\pm$0.056 &-17.99$\pm$0.166 &-18.16$\pm$0.103 &-17.48$\pm$0.202 \\
241275 &60.8 &110.0$\pm$11.46&  34.2$\pm$2.10 &-16.65$\pm$0.135 &-16.55$\pm$0.168 &-16.56$\pm$0.156 &-16.16$\pm$0.167 \\
242042 &69.9 &132.1$\pm$ 2.13 & 33.0$\pm$2.30 &-16.78$\pm$0.15 &--  &--  &--  \\
242246 &63.3 &113.1$\pm$10.08&  61.5$\pm$2.30 &-17.11$\pm$0.086 &-16.50$\pm$0.177 &-16.17$\pm$0.155 &-16.65$\pm$0.125 \\
242248 &61.8 & 74.9$\pm$ 3.41 &102.9$\pm$4.30 &-19.23$\pm$0.09 &--  &--  &--  \\
242284 &68.1 &142.3$\pm$ 2.16&  96.9$\pm$2.30 &-18.56$\pm$0.057 &-17.95$\pm$0.155 &-18.10$\pm$0.068 &-18.10$\pm$0.091 \\
242296 &63.4 &120.8$\pm$ 3.35&  64.3$\pm$2.30 &-18.52$\pm$0.080 &-18.31$\pm$0.135 &-18.36$\pm$0.103 &-18.07$\pm$0.128 \\
248896 &58.0 & 86.1$\pm$ 9.43&  83.2$\pm$2.30 &-16.84$\pm$0.069 &-16.87$\pm$0.189 &-16.78$\pm$0.144 &-16.77$\pm$0.152 \\
248914 &56.1 &110.8$\pm$ 8.43&  62.2$\pm$2.30 &-18.33$\pm$0.082 &-17.86$\pm$0.121 &-18.04$\pm$0.087 &-17.87$\pm$0.102 \\
248970 &58.3 &161.1$\pm$ 8.23 &133.6$\pm$2.20 &-18.81$\pm$0.04 &--  &--  &--  \\
249100 &61.5 &190.1$\pm$ 3.41& 165.1$\pm$2.30 &-19.98$\pm$0.036 &-19.90$\pm$0.077 &-19.68$\pm$0.071 &-19.61$\pm$0.072 \\
249304 &61.3 &115.2$\pm$ 6.84& 103.0$\pm$2.30 &-17.91$\pm$0.056 &-16.99$\pm$0.285 &-18.24$\pm$0.079 &-16.55$\pm$0.340 \\
252891 &60.0 & 93.5$\pm$12.70 & 31.0$\pm$2.30 &-16.79$\pm$0.16 &--  &--  &--  \\
257954 &70.7 &117.6$\pm$ 7.42 & 76.7$\pm$2.30 &-16.28$\pm$0.08 &--  &--  &--  \\
258263 &65.9 &107.3$\pm$ 5.48& 157.5$\pm$2.30 &-17.40$\pm$0.056 &-17.62$\pm$0.393 &-17.89$\pm$0.199 &-17.49$\pm$0.319 \\
258268 &63.5 & 78.2$\pm$ 8.94& 156.4$\pm$2.20 &-18.48$\pm$0.045 &-18.28$\pm$0.245 &-18.28$\pm$0.181 &-18.13$\pm$0.269 \\
258407 &69.1 &271.9$\pm$ 5.35& 150.5$\pm$2.30 &-19.21$\pm$0.044 &-18.95$\pm$0.117 &-18.81$\pm$0.086 &-18.76$\pm$0.100 \\
263839 &63.0 &190.8$\pm$ 7.86& 139.4$\pm$2.30 &-19.86$\pm$0.041 &-19.50$\pm$0.106 &-19.07$\pm$0.168 &-19.11$\pm$0.142 \\
267948 &60.7 &104.4$\pm$ 9.17 & 74.8$\pm$2.30 &-19.00$\pm$0.07 &--  &--  & -- \\
267968 &61.3 &115.1$\pm$ 6.84 &150.7$\pm$2.20 &-18.76$\pm$0.04 &--  &--  &--  \\
268028 &71.1 &199.8$\pm$ 5.29 &104.3$\pm$2.30 &-18.89$\pm$0.05 &--  &--  & -- \\
268107 &66.4 &201.9$\pm$62.22 &191.0$\pm$2.30 &-20.39$\pm$0.03 &--  & -- & -- \\
268151 &56.2 &121.6$\pm$ 3.61 & 76.5$\pm$2.20 &-18.00$\pm$0.07 &--  &--  &--  \\
268199 &71.7 &196.9$\pm$ 6.32 & 70.4$\pm$2.20 &-18.54$\pm$0.07 &--  &--  &--  \\
320466 &66.0 &116.0$\pm$ 2.19 & 43.3$\pm$2.30 &-16.94$\pm$0.12 &--  &--  &--  \\
321385 &62.0 & 84.9$\pm$ 9.06 &102.2$\pm$4.40 &-17.98$\pm$0.10 &--  &--  &--  \\
321490 &71.3 &167.8$\pm$22.17 & 95.0$\pm$2.20 &-18.10$\pm$0.06 &--  &--  &--  \\
332328 &58.1 & 87.2$\pm$ 7.07 & 68.1$\pm$2.10 &-16.47$\pm$0.07 &--  &--  & -- \\
332861 &65.7 &149.2$\pm$21.95& 107.5$\pm$2.20 &-18.78$\pm$0.051 &-18.40$\pm$0.128 &-18.81$\pm$0.078 &-18.52$\pm$0.096 \\
333224 &73.9 &167.6$\pm$ 5.21 &105.4$\pm$2.30 &-18.38$\pm$0.06 & -- & -- &--  \\
714453 &68.5 &248.2$\pm$ 5.37 &151.6$\pm$2.30 &-20.85$\pm$0.04 & -- &  --& -- \\
716496 &73.7 &121.9$\pm$ 4.17 & 28.6$\pm$2.30 &-15.89$\pm$0.18 & -- &--  &--  \\
727027 &59.0 &192.4$\pm$18.66 &195.3$\pm$2.20 &-20.16$\pm$0.03 & -- &--  & -- \\
731695 &55.6 &121.2$\pm$ 6.06 &102.9$\pm$2.30 &-18.50$\pm$0.05 & -- &--  &--  \\
731805 &72.2 &121.8$\pm$ 6.30 & 54.4$\pm$2.50 &-17.06$\pm$0.10 &--  &--  &--  \\
732178 &56.9 &185.0$\pm$ 4.77 & 96.1$\pm$2.30 &-17.94$\pm$0.06 &--  &--  & -- \\
732274 &54.8 &138.3$\pm$ 8.57&  75.1$\pm$2.30 &-18.73$\pm$0.068 &-18.08$\pm$0.130 &-17.78$\pm$0.115 &-17.82$\pm$0.117 \\
732488 &54.9 &180.9$\pm$29.33&  97.9$\pm$2.30 &-18.40$\pm$0.055 &-18.48$\pm$0.117 &-18.75$\pm$0.076 &-17.83$\pm$0.160 \\
732593 &67.1 &146.6$\pm$ 2.17&  42.9$\pm$2.40 &-16.88$\pm$0.124 &-16.31$\pm$0.207 &-16.77$\pm$0.140 &-16.52$\pm$0.154 \\
732937 &60.0 &170.9$\pm$ 8.08 & 71.4$\pm$2.20 &-17.79$\pm$0.07 &--  &--  &--  \\
733774 &56.8 &163.8$\pm$10.76 &151.4$\pm$2.20 &-19.42$\pm$0.04 &--  &--  &--  \\
748757 &76.7 &136.7$\pm$ 2.06 & 56.1$\pm$2.40 &-17.60$\pm$0.10 & -- & -- & -- \\
748786 &67.5 &161.3$\pm$ 5.41&  77.7$\pm$2.20 &-17.49$\pm$0.070 &-17.00$\pm$0.281 &-17.29$\pm$0.163 &-16.17$\pm$0.459 \\
748819 &74.5 & 55.0$\pm$ 2.08&  35.0$\pm$2.20 &-15.28$\pm$0.141 &-16.55$\pm$0.185 &-15.88$\pm$0.186 &-14.09$\pm$0.844 \\
749209 &69.0 &155.3$\pm$25.71 &103.4$\pm$2.20 &-17.90$\pm$0.06 & -- & -- & -- \\
749230 &76.5 &130.6$\pm$10.28 & 60.4$\pm$2.30 &-16.87$\pm$0.09 & -- & -- &--  \\
749264 &58.3 &170.3$\pm$ 7.05& 103.4$\pm$2.30 &-18.29$\pm$0.053 &-18.31$\pm$0.086 &-18.46$\pm$0.104 &-18.27$\pm$0.107 \\
749287 &71.5 &157.1$\pm$ 7.38 & 98.0$\pm$2.10 &-17.77$\pm$0.06 & -- &--  & -- \\
749324 &62.6 & 63.1$\pm$ 5.63 & 24.7$\pm$2.20 &-15.37$\pm$0.19 &--  & -- &--  \\
749362 &60.3 &201.4$\pm$ 9.21& 130.0$\pm$2.40 &-18.63$\pm$0.047 &-18.21$\pm$0.157 &-18.80$\pm$0.107 &-17.97$\pm$0.189 \\
749363 &61.5 &117.2$\pm$ 5.69&  85.0$\pm$2.30 &-17.99$\pm$0.063 &-17.91$\pm$0.116 &-18.13$\pm$0.101 &-18.02$\pm$0.124 \\
749368 &63.6 & 56.9$\pm$ 3.35 & 64.6$\pm$2.30 &-15.98$\pm$0.09 &--  & -- &--  \\
749380 &76.2 &144.2$\pm$ 4.12&  86.0$\pm$2.40 &-17.74$\pm$0.069 &-17.29$\pm$0.114 &-17.54$\pm$0.135 &-17.08$\pm$0.174 \\
749383 &56.7 &116.0$\pm$ 7.18& 135.4$\pm$2.30 &-18.05$\pm$0.048 &-17.95$\pm$0.171 &-17.88$\pm$0.279 &-17.97$\pm$0.264 \\
749405 &72.5 &154.1$\pm$10.48 & 56.7$\pm$2.20 &-16.75$\pm$0.09 & -- &--  &--  \\
749423 &60.5 &121.8$\pm$13.79 & 95.3$\pm$2.10 &-18.34$\pm$0.05 & -- &  --&  --\\
749438 &55.8 &128.2$\pm$ 8.46 & 53.4$\pm$2.30 &-16.07$\pm$0.10 &--  & -- &--  \\
749452 &59.5 & 92.9$\pm$ 8.12 & 40.8$\pm$2.40 &-15.32$\pm$0.13 & -- &--  & -- \\
749464 &74.9 &133.6$\pm$10.36&  95.8$\pm$2.30 &-17.44$\pm$0.068 &-15.37$\pm$1.447 &-16.51$\pm$0.407 &-16.67$\pm$0.321 \\
749475 &62.0 &132.5$\pm$ 9.06 &179.5$\pm$2.20 &-18.35$\pm$0.05 &--  & -- & -- \\
  \enddata
\end{deluxetable*}
\end{longrotatetable}
To investigate the TFr, the W50 of the H{\sc{i}} line should be first of all corrected 
for inclination effect. For the subsample galaxies, we have the minor-to-major axis ratios, b/a 
(q in Table ~\ref{tab:cata}), which were estimated by us fitting the $g$-band galaxy images 
with a single exponential profile model. As galaxies in this subsample are visually round,
the b/a ratios derived from a single exponential model fitting for this subsample galaxies
are more appropriate.

Using the b/a ratios (q), we corrected the width of the H{\sc{i}} line width, W50, for inclination effect via Equation (2) below. The inclination angle,$i$, could be calculated using the formula proposed by\citet{Hubble1926} which is also listed as Equation (3) below in this paper. 
The intrinsic axis ratio,q$_{0}$, in Equation (3) was set to be 0.2 \citep{Masters2008} 
to mitigate the effect of the bulge on the most edge-on galaxies.
\begin{eqnarray}
   W50_{cor} &=& W50/sin(i)\\
   cos(i)^{2}&=& \frac{q^{2}-q_{0}^{2}}{1-q_{0}^{2}}
\end{eqnarray} 

Additionally, the W50 has already been corrected for instrumental broadening by the ALFALFA team \citep{Haynes2011}. 
Although the TFr was originally discovered in the optical bands, 
it was later proved to be even tighter in the near-infrared (NIR) bands (see more in Section ~\ref{sec:intro}). Therefore, we would study the TFrs  in the optical $B$, $g$ and $r$ bands and 
the NIR $J$, $H$ and $K$ bands for the LSBG subsample.
 
\subsection{Optical TFr}
In this subsection, we study the optical $B$-, $g$- and $r$-band TFrs for the LSBG subsample. 
First of all, we converted the $B$-band magnitudes (measured by ourselves in \citet{Du2015}
and listed in Table ~\ref{tab:cata}) to absolute magnitudes, M$_{B}$, for the subsample galaxies.
The distances used for conversion are directly from the newly released ALFALFA Extragalactic 
H{\sc{i}} source catalogue \citep{Haynes2018}. Compared with the $\alpha$.40 catalog \citep{Haynes2011}, 
the newly released ALFALFA catalog has provided uncertainties on distance. So, the errors on $B$-band 
absolute magnitude were propagated from the errors on $B$-band magnitude and the distance 
(provided by the ALFALFA catalogue), based on the error transfer formula of mathematical statistics. 
In order to compare the TFr of our LSBGs with the TFrs of other available LSBG samples,
such as the LSBG sample of \citet{Zwaan1995}, in $B$ band, we adopted the same prescription 
as was used in \citet{Zwaan1995} to correct the M$_{B}$ for Galactic and internal extinction to 
face-on orientation.This prescription (Equation (4) below) for extinction correction was proposed
by \citet{Tully1985}.

\begin{eqnarray}
  M(B)_{corr}=M_{B}+2.5log[f(1+e^{-\tau sec(i)})+(1-2f)(\frac{1-e^{-\tau sec(i)}}{\tau sec(i)})]
\end{eqnarray} 

where $\tau$ is the optical depth, and $f$ is the fraction of the light that is unobscured by the dust layer.  
 $\tau \sim$ 0.55 and f $\sim$0.25 were adopted in most publications\citep{Tully1985,Zwaan1995}.

\begin{figure}[ht!]
\epsscale{1.2}
\plotone{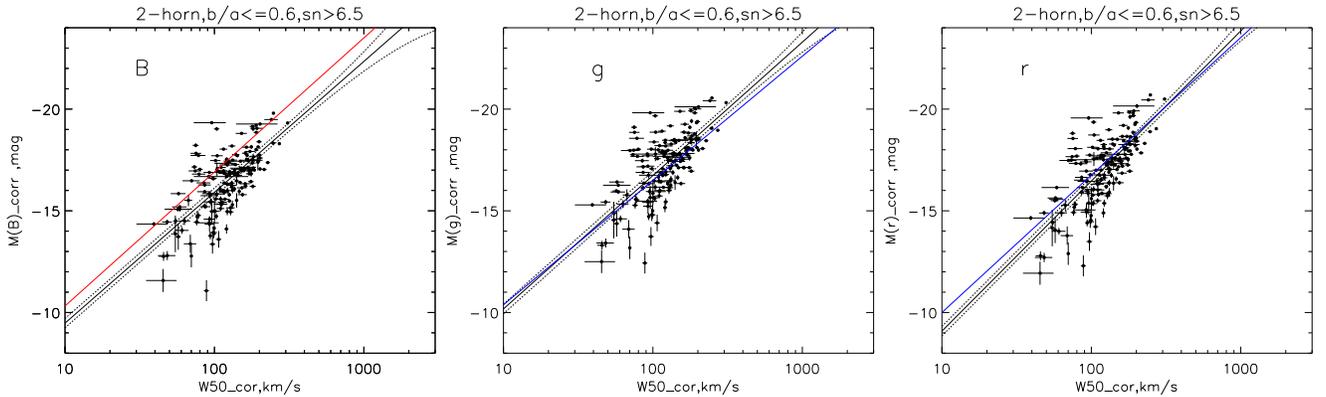}
\caption{Tully-Fisher diagrams in $B$(left), $g$(middle) and $r$(right) bands for our LSBG subsample.  
We show the linear fit (black solid line) and the corresponding 95$\%$ confidence bands (black dashed hyperbolae) of the fit for the LSBG subsample. The error bars along both coordinates for each of our data points are shown. The TFr line followed by the Broeil sample of HSB spiral galaxies 
and also followed by a small sample of LSBGs from \citet{Zwaan1995}
is overlaid as the red line in Figure ~\ref{fig:fig_tf}(a).
Besides, the $g$- and $r$-band TFr lines 
which are followed by a sample of 32 large and relatively nearby galaxies
from \citet{Ponomareva2017} are also over-plotted as blue lines 
in Figures ~\ref{fig:fig_tf}(b) and (c).\label{fig:fig_tf}}
\end{figure}

In Figure ~\ref{fig:fig_tf} (a), we plot our LSBG subsample (black dots) in the plane of corrected H{\sc{i}} 
line width v.s. corrected absolute magnitude (TF plane) in $B$ band.The linear regression line (black solid line)
of the TFr and the corresponding 95\% confidence bands (black dashed hyperbolae) of the regression line
are over-plotted. The function we used for regressing the TFr line is the MPFITEXY function in IDL. 
This function takes into account errors in both axises and performs an inverse least squares regression 
which can resolve the Malmquist bias \citep{Willick1994} that affects the slope of the TFr.
Furthermore, in order to estimate 
the confidence intervals for the coefficients of the regression lines, we performed pair bootstrap resampling for 
1000 times. Quantitatively, the vertical scatter and the tightness of the TFr fit are also measured following the 
method of \citet{Ponomareva2017}. The coefficients, confidence intervals, scatter and tightness of the TFr
fit are also quantitatively listed in Table~\ref{tab:tf_reg}.

For comparison with the TFr of the LSBG sample of \citet{Zwaan1995}, we over-plotted the $B$-band TFr line 
which is followed by the LSBG sample of \citet{Zwaan1995} as the red line in Figure ~\ref{fig:fig_tf}(a).
Apparently in Figure ~\ref{fig:fig_tf}(a), the slope of the TFr line (black solid line) for our LSBG subsample
seem to be well consistent with the slope of the TFr line (red solid line) for the LSBG sample of \citet{Zwaan1995}. 
Importantly, this blue TFr line was in \citet{Zwaan1995} to be well consistent with the TFr of the High Surface 
Brightness Galaxies (HSBGs) compiled by \citet{Broeils1992}. Therefore, we could indirectly deduce that 
the slope of the TFr line for our LSBG subsample is also in good agreement with the slope of the TFr line for 
HSBGs, albeit our LSBG sample has a larger scatter than the HSBGs ($\sigma$=1.14~mag v.s. 0.77~mag) 
in $B$ band. 
 
Besides, \citet{Ponomareva2017} has reported TFrs for 32 large and relatively nearby galaxies which are generally HSBGs 
in $g$ and $r$ bands. 
So we also expect to compare the TFr of our LSBG subsample 
with the TFr of the HSBGs from \citet{Ponomareva2017} in $g$ and $r$ bands. 
We made the Galactic and internal extinction correction for the absolute 
magnitudes in $g$ and $r$ bands for the LSBG subsample, using the same 
prescription as was used by \citet{Ponomareva2017}. 
According to \citet{Ponomareva2017b},
the Galactic extinction correction was obtained using the ``Galactic dust reddening and extinction'' tool
provided by the NASA/IPAC infrared science archive which estimates the Galactic dust extinction
from \citet{Schlafly2011}. For $g$-band magnitude, the internal extinction correction, $A_{g}^{i}$, was
assumed to be 0.6~mag which is an average internal extinction of spiral galaxies from face-on to edge-on.
For $r$-band magnitude, the internal extinction correction, $A_{r}^{i}$, can be expressed as: 
\begin{eqnarray}
   A_{r}^{i} &=& \gamma_{r} log(a/b)\\
   \gamma_{r} &=& 1.15 + 1.88(logW_{50}^{i} - 2.5)
\end{eqnarray} 
where $a/b$ is the major-to-minor axis ratio of the galaxy and $\gamma_{r}$ is an extinction amplitude parameter 
which was calibrated by \citet{Tully1998} as a function of the width of the H{\sc{i}} line profile corrected for inclination.
After correcting the Galactic and internal extinction for the absolute magnitude in $g$ and $r$ bands 
using the prescription described above which was used in \citet{Ponomareva2017},  
we plot our LSBG subsample (black filled circles) and the regression TFr lines (black solid lines fitted by MPFITEXY function in IDL)
in $g$- and $r$-band TFr planes in Figures ~\ref{fig:fig_tf} (b) and (c).
The confidence bands of the fit were shown as black dashed hyperbolae. 
For comparison, the $g$- and $r$-band TFr lines for the \citet{Ponomareva2017} sample (HSBGs) 
were also over-plotted as blue lines in Figures ~\ref{fig:fig_tf} (b) and (c).
The coefficients, confidence intervals, scatter and tightness of the TFr fit in $g$ and $r$ bands are also
listed in Table ~\ref{tab:tf_reg}.
Within the uncertainties of the value of the TFr slope (listed in Table ~\ref{tab:tf_reg}), 
the TFr slopes for our LSBG subsample are
generally in agreement with the TFr slopes for the HSBGs from \citet{Ponomareva2017}
in $g$ and $r$ bands, 
although the scatter ($\sigma \sim$1.17~mag) and tightness ($\sim$ 0.17~mag) of the TFr fit 
for our LSBGs is not so good as the scatter ($\sigma$=0.27) and tightness (0.068~mag) of the TFr fit 
for the HSBGs from \citet{Ponomareva2017}.

Conclusively, in the optical $B$, $g$ and $r$ bands, 
the TFr lines for our LSBG subsample agree with the TFr lines for
HSBGs in slope.This result is consistent with the previous optical TFr results for LSBGs
from \citet{Zwaan1995} and \citet{Sprayberry1995}, 
which confirmed that the LSBGs follow the same TFr as the normal spiral galaxies (HSBGs) 
in the optical bands. However, compared with the HSBGs in Figure ~\ref{fig:fig_tf}, our LSBGs 
have relatively larger scatter and are less tighter around our TFr lines,
which might be due to the variations in mass-to-light ratio among our LSBGs.
 
\subsection{NIR TFr}
In the subsample of 173 LSBGs which are appropriate for studying TFr,
99 galaxies have been observed in $J$, $H$ and $K$ bands 
by the Large Area Survey (LAS) of UKIRT Infrared Deep Sky Survey (UKIDSS).
In this subsection, we investigate the TFrs for these 99 LSBGs in $J$, $H$ and $K$ bands.

First of all, we subtracted the sky background for NIR band images of galaxies
in $J$, $H$ and $K$ bands, using the same method as described in \citet{Zheng1999},
\citet{Wu2002} and \citet{Du2015}. 
Then, the magnitudes were measured on the sky-subtracted images
within the AUTO aperture by SExtractor. The measured magnitudes were then 
corrected for Galactic extinction using the prescription of \citet{Schlafly2011}.
The internal extinction of LSBGs is very difficult to estimate.
\citet{Tully1998} found that the internal extinction was unmeasurably 
small for faint galaxies with  M$_{B} \ge$-17~mag which is the case for most of 
the galaxies in our LSBG subsample (see Figure ~\ref{fig:fig_tf}(a)).
Most LSBGs are thought to be poor in dust based on the statistically inferred 
low internal extinction (e.g., \citet{Tully1998}; \citet{Masters2008}).
Therefore, we did not make any internal extinction correction for 
the NIR-band magnitudes of our LSBG subsample so as to avoid
large uncertainties that might be introduced in the magnitudes due to internal correction.
The details on data reduction including sky subtraction and surface photometry
in the NIR bands will be reported in our forthcoming paper 
which will be on spectral energy distributions (SED) of our entire LSBG sample.
Then, the NIR-band magnitudes were converted to absolute magnitude by using
the distance from the ALFALFA catalogue.

\begin{figure}[ht!]
\epsscale{1.2}
\plotone{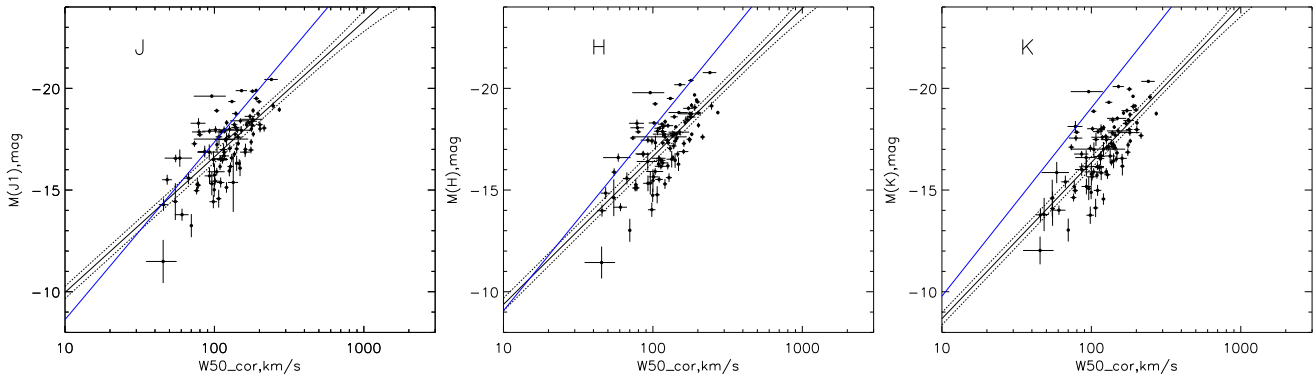}
\caption{Tully-Fisher relation diagrams in the NIR $J$ (left), $H$ (middle) and $K$ (right) bands 
for the LSBG subsample (black dots), based on magnitudes without internal extinction correction.
We show the regression TFr lines (black solid lines) and 
 the corresponding 95$\%$ confidence bands (black dashed hyperbolic curves) 
 of the black regression lines for the LSBG subsample in each band.
 The error bars along both coordinates for our data points are also shown.
 For a comparison, the TFr lines 
which are followed by the sample of 32 large and relatively nearby galaxies
from \citet{Ponomareva2017} in 2MASS $J$- , $H$- and $Ks$ bands
 are also overplotted as blue lines.\label{fig:fig_tf_nir}}
\end{figure}

Based on the corrected W50 of H{\sc{i}} and the absolute magnitudes without internal extinction correction, 
we show these 99 LSBGs (black dots) in the TFr planes in $J$, $H$ and $K$ bands, respectively, 
and the corresponding regression TFr lines (black solid lines; by the MPFITEXY function in IDL)
in Figure ~\ref{fig:fig_tf_nir}. The coefficients, confidence bands, scatter and tightness of the regression lines 
(black lines in Figure ~\ref{fig:fig_tf_nir}) for $J$, $H$ and $K$ bands were tabulated in the middle three lines 
in Table ~\ref{tab:tf_reg}. For comparison, we overplotted the TFr lines of the three NIR bands
for the HSBGs of \citet{Ponomareva2017} as the blue solid lines in Figure ~\ref{fig:fig_tf_nir}. 

\begin{figure}[ht!]
\epsscale{1.2}
\plotone{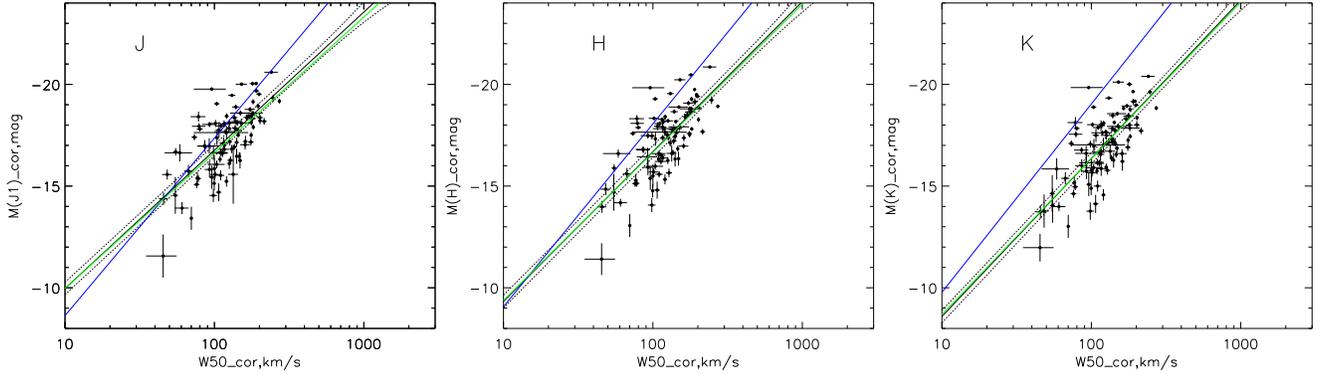}
\caption{Tully-Fisher diagrams in the NIR $J$ (left), $H$ (middle) and $K$ (right) bands for the LSBG subsample (black dots), based on the magnitudes corrected for internal extinction. The error bars along both coordinates for the data points are shown. The black solid lines are the linear regression lines (by the MPFITEXY function in IDL) based on the corrected magnitudes and the black dashed hyperbolic curves show the corresponding 95$\%$ confidence bands of the black regression lines. The blue solid lines represent the TFr lines for the \citet{Ponomareva2017} sample which is composed of 32 large and relatively nearby galaxies in $J$- , $H$- and $Ks$ bands based on the 2MASS data.
For comparison, the TFr lines which was previously regressed based on the magnitudes without 
internal extinction were also overplotted as the green lines (Note: The green lines are exactly the black
solid lines in Figure ~\ref{fig:fig_tf_nir}). \label{fig:fig_tf_nir2}}
\end{figure}

Compared with optical bands, the slopes of the NIR-band TFr lines (black solid lines)
for our LSBG subsample do not appear to be well consistent with the slopes of the NIR-band
TFr lines (blue solid lines) for the HSBGs of \citet{Ponomareva2017}.
In each panel of Figure ~\ref{fig:fig_tf_nir}, the offset in magnitude between the two TFr lines 
for LSBGs (black solid line) and HSBGs (blue solid line)
increases with the line width.
We think this might be caused by the effect of internal extinction correction for magntiudes. 
As LSBGs are well known to have little 
internal extinction, especially in the NIR bands,
we did not do any internal extinction correction 
for magnitudes in $J$, $H$ and $K$ bands for our LSBGs
to avoid large uncertainties that might be introduced into the magnitudes.  
However, the internal extinction is a significant effect on magnitudes for HSBGs, 
so \citet{Ponomareva2017} corrected the magnitudes  in $J$, $H$ and $K$ bands
for internal extinction for their HSBGs. 
The prescription they used for correcting
internal extinction was from \citet{Tully1998} and \citet{Sorce2012a} 
\citep{Ponomareva2017b}, which calibrated
the internal extinction value as an increasing function of the width of 
the H{\sc{i}} line. This may account for the increasing magnitude offset between
the TFr lines of our LSBGs (black solid lines in Figure ~\ref{fig:fig_tf_nir}) and 
the HSBGs (blue solid lines in Figure ~\ref{fig:fig_tf_nir}) from \citet{Ponomareva2017} 
with increasing line width.
To check the effect of the internal extinction correction, 
we attempted to correct the internal extinction for $J$-, $H$- and $K$-band absolute magnitudes
for our LSBG subsample, using the same prescription as was used by \citet{Ponomareva2017} and 
detailedly listed in \citet{Ponomareva2017b}. 
Based on the corrected absolute magnitudes,
we further studied the NIR-band TFr regressions (black solid lines) in Figure ~\ref{fig:fig_tf_nir2}.
The coefficients of the new TFr lines based on the corrected magnitudes in
$J$, $H$ and $K$ bands are tabulated in the last three lines in Table ~\ref{tab:tf_reg}.
In order to better compare the NIR TFr lines based on magnitudes before and after internal extinction correction, the TFr lines which was previously regressed before the internal extinction correction for NIR-band magnitudes 
were overplotted as the green lines in Figure ~\ref{fig:fig_tf_nir2} which are exactly the black
solid lines in Figure ~\ref{fig:fig_tf_nir}. By comparison, no big difference was found
between the TFr lines regressed before and after correcting the internal extinction for the magnitudes
in $J$, $H$ and $K$ bands, suggesting that the internal extinction correction is
not an important factor which caused the difference between the TFr lines of LSBGs and HSBGs. 
This result further strengthens that the the dust internal extinction is very little for LSBGs.

\begin{deluxetable*}{lcccccc}
\tablecaption{Regression coefficients for our LSBG sample in Tully-Fisher diagram in optical $B$, $g$, $r$, NIR $J$, $H$, and $K$ bands, 
and for different luminosity types and morphological types in $B$ band. \label{tab:tf_reg} }
\tablewidth{750pt}
\tablehead{
 \colhead{Band} & \colhead{k} & \colhead{b}  &\colhead{95$\%$ CI for $k$} & \colhead{95$\%$ CI for $b$} & \colhead{scatter} &\colhead{tightness} 
} 
\startdata
 $B$  & -6.43$\pm$0.51   &-3.04$\pm$1.07   & [-7.48, -5.41]   & [-5.26, -0.77] & 1.14  &0.18\\
 \hline
 $g$  & -6.56$\pm$0.53   &-3.62$\pm$1.11    & [-7.55, -5.55]   & [-5.82, -1.47] &1.18  & 0.18\\
 \hline
 $r$   & -7.38$\pm$0.54   &-1.70$\pm$1.12    & [-8.45, -6.35]   & [-3.88, 0.61] &1.17   &0.16\\
 \hline
 \hline
 $J$  & -6.67$\pm$0.72   &-3.30$\pm$1.50    &[-8.12,-5.18]     &[-6.55,-0.17]     & 1.17 &0.17\\
 \hline
$H$  & -7.28$\pm$0.74   & -2.10$\pm$1.55   & [-8.71, -5.76]   & [-5.34, 0.93]   &1.19  & 0.16\\
  \hline
$K$   & -7.68$\pm$0.74  & -1.00$\pm$1.55   & [-9.08, -6.35]    &[-4.05, 2.01]   &1.16  &0.15\\
\hline
\hline
 $J$  & -6.79$\pm$0.71   &-3.17$\pm$1.49    &[-8.29,-5.25]     &[-6.56, 0.11]  & 1.17  &0.17\\
 \hline
$H$  & -7.37$\pm$0.74   & -1.96$\pm$1.55   & [-8.65, -5.83]   & [-5.20, 0.80]   &1.19  & 0.16\\
  \hline
$K$   & -7.77$\pm$0.74  & -0.81$\pm$1.55   & [-9.03, -6.41]    &[-3.69, 1.84]   &1.16  &0.15\\
 \enddata
 \tablecomments{The regression line can be expressed as  $y$=k$x$+b, 
 where $x$ represents W50$_{corr}$ in logarithm, $y$ represents absolute magnitude in $B$, $g$, $r$, 
 $J$, $H$, or $K$ bands,$k$ is the slope, and $b$ is the intercept.CI represents Confidence Interval.}
\end{deluxetable*}

Then, we have to find other possible reasons which might cause the difference between the TFr lines
of our LSBGs and the HSBGs from \citet{Ponomareva2017}.
Compared with the bright disk of HSBGs, the outer disk of LSBGs are 
fainter and thus more difficult to be detected in NIR bands than HSBGs. 
Generally, the surface brightness profiles of LSBGs 
are systematically 3.5~mag fainter than normal HSBGs in $J$ band
\citep{Galaz2002}. 
Therefore, with the shallow surface brightness limiting of UKIDSS,
the disk of the extended LSBG could only be detected out to a small radius in NIR bands,
compared to the outmost edge of the optical disk.
However, for HSBGs in \citet{2017MNRAS...469...2387}, 
even although their NIR-band magnitudes
are measured on 2MASS images which are even shallower than UKIDSS, 
the HSBG disks can still be detected out to 
a larger radius close to the outmost edge of the optical disk.
So this would cause more serious underestimation
of the total disk luminosity for LSBGs than for HSBGs in NIR bands,
which can then cause the magnitude offset between the TFr lines for LSBGs
and HSBGs. This might be one possible reason for the difference between
the TFr lines of LSBGs and HSBGs in NIR bands, and more possible reasons
needs to be explored in the further work.

\section{Discussion}\label{sec:discuss}
\subsection{Strength and weakness of our LSBG sample}
As for strength, firstly, we need the redshift information to correct the cosmological dimming
of the central surface brightnesses of galaxies.
However, SDSS has spectroscopic redshifts only for those galaxies which are brighter than 17.77~mag in $r$ band and called the Main Galaxy Sample of the SDSS survey. 
As most of LSBGs are fainter in optical band,
it means that a large fraction of true LSBGs could not be defined by only using the optical survey of SDSS
due to the lack of accurate redshifts from spectroscopy. 
For example, as shown in Figure ~\ref{fig:fig_para} (b), 
31$\%$ of galaxies in our LSBG sample are fainter than 17.77~mag in $r$ band.
So if we were selecting our LSBG sample only on the basis of the optical SDSS survey,
those 31$\%$ faint LSBGs would have been lost from our final sample.
However, our LSBG sample is an H{\sc{i}}-selected sample 
the radial velocities of which, especially the faint LSBGs,
are all available from the $\alpha$.40 H{\sc{i}} catalogue.
So our H{\sc{i}}-selected LSBG sample does not limited by the magnitude of the SDSS Main Galaxy Sample.
Instead, it includes more faint LSBGs (31$\%$ fainter than $r$=17.77~mag)
which could reach down to the surface brightness depth of the SDSS photometric survey
but could not be confirmed by only using the SDSS data due to lack of the spectroscopic redshift.

Secondly, LSBG selections are very sensitive to the subtraction of sky background of images.
However, as explained detailedly in Section 3.1 in \citet{Du2015}, the
SDSS photometric pipeline has a high risk of considering the large extended outskirts of the objects as part of the sky background
, and thus overestimated the sky background of galaxies which further leads to the underestimation of the luminosity of galaxies.
The underestimation in luminosity is $\sim$0.2~mag for bright galaxies with extend outskirts and is up to $\sim$0.5~mag for LSBGs. 
Therefore, we reconstructed the sky background map for all galaxy images, using a more precise method 
which were described and testified in detail in Section 3.1 in \citet{Du2015}.
Thus, we corrected the underestimation of luminosity for galaxies in our LSBG sample.

As for the weakness, on one hand, our LSBG sample was selected basing on the central disk surface brightness 
which was derived by fitting the galaxy image with a single exponential profile model with Galfit.
So for galaxies which have light concentration or bulges in the centers, the central disk surface brightnesses of the disk must have been 
overestimated (bias to brighter values). However, we made some tests in \citet{Du2015} 
by fitting the LSBGs with two components (Sersic+Exponential) and found that the overestimation for the disk central surface brightness were systematically
very small and less than 0.2~mag/arcsec$^{2}$. Additionally, the part of bulge-dominated galaxies (only 5.8$\%$) is not dominant in our LSBG sample. 
In the case of overestimation, the true central disk surface brightnesses for those galaxies should be even fainter than currently, so these galaxies currently in our LSBG
sample should still be members of the LSBG sample. However, it is poor in bulge-dominated LSB galaxies due to our fitting method of only using 
a single exponential profile precluding the detection of LSBGs with bulges. But this fitting method is fast and valid for the fast majority of the LSBGs.
On the other hand, as selected from the H{\sc{i}} survey,
our LSBG sample is composed of more gas-rich galaxies, 
but should have been deficient in the true gas-poor LSBGs.
If there was no such effects explained above, our LSBG sample would on one hand include more galaxies with large central light concentration or bulges.
These galaxies would have large rotational velocities and bright luminosities so that our LSBG sample would be enlarged and stretched to the ends of 
larger W50 and brighter absolute magnitude in the TF plane in Section 5. On the other hand,  our LSBG sample would have more gas-poor galaxies 
which would be more likely to enlarge or stretch our sample to
the end of smaller W50 and lower absolute magnitude in the TF plane in Section 5.
With those possibly-added LSBGs in the TF plane, the TFr fit in Section 5 would probably be better constrained.

\subsection{Mass-to-light ratios}\label{sec:ml}
Theoretically for a spiral galaxy, the mass $M$ is proportional to $V_{max}^{2}h$ 
and the total luminosity $L$ is proportional to $\Sigma_{0}h$,
where $V_{max}$ is the maximal rotational velocity, $h$ is the disk scale length
and $\Sigma_{0}$ is the central surface brightness. From these two relations,
a simple relation between the luminosity ($L$), 
the maximal rotational velocity ($V_{max}$), 
central surface brightness ($\Sigma_{0}$), 
and mass-to-light ratio ($M/L$) follows that
 \begin{equation}
   L \propto \frac{V_{max}^{4}}{\Sigma_{0}(M/L)^{2}}
 \end{equation}
The slope $\alpha$ of the TFr in the form of $L \propto V_{max}^{\alpha}$ 
generally changes with the observed band and gradually 
increases from $\alpha\sim$~2.5 in the optical $B$ band 
to $\alpha\sim$~4 in the NIR $K$ band (e.g.[]\citet{Aaronson1983,Giovanelli1997,Courteau2007}).
In Equation (), $L \propto V_{max}^{4}$ is recognized as
the universal TFr.

As we concluded in Section 5 that
our LSBG subsample follows a normal TFr that is followed by the normal HSBGs, 
we could attempt to explore the clues for the total mass-to-light ratio $M/L$
of these LSBGs by statistically comparing with the $M/L$ of the Broeil's HSBGs below, 
following the comparison method of \citet{Zwaan1995}.
In Figure ~\ref{fig:fig_tf} (a), 
the $B$-band slope of TFr (black solid line) for our LSBG subsample
is -6.43$\pm$0.51 (Table ~\ref{tab:tf_reg}), which is in good agreement with the 
$B$-band slope of TFr (red line) for the HSB spiral galaxy sample 
of \citet{Broeils1992} (hereafter Broeils spiral sample), which is found to be -6.59.
The intercept of the black line and red line are, respectively,
 -3.04$\pm$1.07  (Table ~\ref{tab:tf_reg}) and -3.73$\pm$0.77 for the Broeils spiral sample \citep{O'Neil2000}.
So albeit the consistent slope, there is a systematical offset of $\sim$0.7~mag in absolute magnitude
between the two TFr lines.
This means that our LSBGs are systematically $\sim$0.7~mag fainter than the Broeils HSBGs in $B$ band magnitude, 
which can be expressed as M$_{LSBG}$-M$_{HSBG}$=0.7~mag.
As L$_{LSBG}$/L$_{HSBG}$=10.0$^{-0.4\times(M_{LSBG}-M_{HSBG})}$, the luminosity
of our LSBGs is then systematically $\sim$~0.5 times (=10.0$^{-0.4\times0.7}$) 
the luminosity of the Broeils HSBG sample in $B$ band at a fixed H{\sc{i}} line width.
As for central surface brightness, the mean $B$-band central surface brightness is 
23.1 mag~arcsec$^{-2}$ for our LSBG subsample and 21.6 mag~arcsec$^{-2}$ for the Broeils HSBGs,
so our LSBG subsample is 1.5 $mag$ fainter than, or 1/4 (=10$^{-0.4\times1.5}$) times, 
the typical central surface brightness value of the Broeils HSBGs.
Therefore, according to Equation (7), in order to account for the factors of differences in $\Sigma_{0}$ and $L$ at the fixed line velocity, 
it can be deduced that $M/L$ of our LSBGs must be $\sim$2.8 times of 
the $M/L$ of Broeils HSB spiral galaxy sample.

However,
although the NIR $K$-band slope of $\sim$4 ($L \propto V_{max}^{4}$) is recognized as
 the universal TFr (The slope $k$ is corresponding to $\sim$ 10 
when the luminosity is expressed in absolute magnitudes, $M$,
and the TFr is in the form of $M=-k log(V_{max})+b$).
it should be noticed that the derivation of the TFr 
has been questioned by recent papers 
(e.g.\citet{Courteau2007,Ponomareva2017}), which obtained shallower slopes ($\alpha$<4)
 in agreement with their own data in NIR $K$ band. 
 For our LSBG subsample, we also obtained a shallower TFr.
 So, insisting on using the theoretical Equation (7) which reflects a universal TFr ($L \propto V_{max}^{4}$), 
 we might have overestimated the luminosity, $L$, of our LSBG subsample and thus 
 overestimated the luminosity ratios between our LSBG subsample and the Broeil
HSBG sample at a fixed rotational velocity, and finally underestimated the $M/L$
of our LSBG subsample. So the ratio between the total $M/L$ of our LSBG subsample and
the Broeils HSBG sample should be larger than $\sim$2.8 that was deduced above based on 
a universal TFr of the slope of 4.

\subsection{Disk scale lengths}
Since $M/L \propto V_{max}^{2}h/L$, 
where $h$ is the disk scale length, $M/L$ depends on $h/L$ at a fixed line width.
In \S ~\ref{sec:ml}, using Equation (7),
we derived that the $M/L$ of our LSBG subsample was 
2.8 times the $M/L$ of the Broeils HSB spiral galaxy sample,
and $L$ (luminosity) of our LSBG subsample was 0.5 times the luminosity of 
the Broeils HSB spiral galaxy sample at a fixed line width.
So, the disk scale length of our LSBG subsample should be 1.4 times of 
the disk scale length of the Broeils HSB spiral galaxy sample.

\begin{figure}[ht!]
\epsscale{0.7}
\plotone{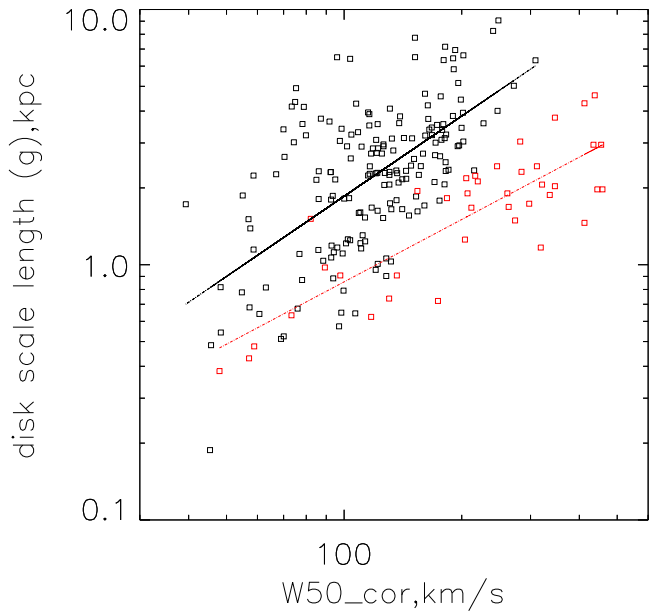}
\caption{Corrected H{\sc{i}} line width vs. disk scale length. 
Black open squares are LSBGs of our subsample and 
the black line is the linear regression for our LSBG subsample. 
Red open squares are spiral galaxies from the Broeils sample 
used in \citet{1995MNRAS...273...L35} and
the red line is the linear regression for Broeils sample.\label{fig:scalelength}}
\end{figure}

In Figure ~\ref{fig:scalelength}, we plot the disk scale length versus the corrected H{\sc{i}} line width
for our LSBG subsample in $g$ band. For comparison, HSBGs of the Broeils sample are over-plotted
in Figure ~\ref{fig:scalelength}. It is worth noting that the disk scale length for HSBGs of the Broeils sample 
in Figure ~\ref{fig:scalelength} are in $B$ band because only the $B$-band the disk scale lengths are available 
for Broeils sample. This would not seriously affect the results about disk scale length later since
galaxy size in the $B$-band image appears very similar to that in $g$ band.  

Apparently in Figure ~\ref{fig:scalelength}, a trend exists between disk scale length and line width for both
the LSBGs and the HSBGs. To make the trend as clear as possible, we made a linear regression 
between line width and disk scale length for both the LSBG subsample (black solid line) and 
the Broeils's HSBG sample (red dashed line). It obviously appears that our LSBG subsample is systematically 
larger than the HSBG sample in disk scale length. 
Observationally, a minor difference between the slopes (1.05 for LSBGs v.s. 0.81 for Broeils's HSBGs) of both lines 
is present, and the offset between the two fitting lines is 0.24~dex at W50 $\sim$ 40~km/s 
(the minimal corrected W50 for our LSBG subsample) 
and 0.45~dex at W50 $\sim$ 300~km/s (the maximal corrected W50 for our LSBG subsample).
The offsets implies that our LSBGs have disk scale lengths that are 1.75(=10$^{0.24}$) times larger 
than the HSBGs of Broeils sample at the minimal corrected W50 and 
2.82(=10$^{0.45}$) times larger than the HSBGs of Broeils sample at the maximal corrected W50.
Therefore, our LSBGs are 1.75$\sim$2.54 times 
larger in size than the HSBGs from the Broeils sample at the fixed line width.
This factor range from observation is slightly higher than the 
factor of 1.4 deduced from the M/L above using Equation (7).
This discrepancy might be because that our LSBG subsample observationally
follows a shallower TFr
so that the M/L of our LSBGs 
has been underestimated by Equation (7) which is based on a universal TFr.

\subsection{Mass surface density}\label{msd}
From a universal TFr, it follows that $L \propto V_{max}^{4}$ and $V_{max}^{4} \propto M^{2}/h^{2}$.
So, we can derive that $L \propto M^{2}/h^{2}$ or that
 \begin{equation}
 \frac{M}{L} \propto \frac{h^{2}}{M} = \frac{1}{\bar{\sigma}}
  \end{equation}
where $h$ is the disk scale length and $\bar{\sigma}$ is the average mass surface density.
This relation declares that galaxies with similar mass surface densities have similar $M/L$ 
while galaxies with higher $M/L$ have lower $\bar{\sigma}$.
As derived in \S ~\ref{sec:ml}, for LSBGs of our subsample,
the $M/L$ is 2.8 times 
the $M/L$ of spiral galaxies of the Broeils sample, 
so the $\bar{\sigma}$ for the subsample of our LSBGs
is $\sim$1/2.8$\simeq$0.4 times
the mass surface density of the spiral galaxies of the Broeils sample.
As discussed in Section 6.2, the $M/L$ might have been underestimated for
our LSBGs by a universal TFr of slope $\sim$ 4, so the $\bar{\sigma}$ might have
been overestimated here for our LSBG subsample.
Any way, this result proves that LSBGs have lower mass surface densities than 
the HSBGs (the Broeils spiral sample) with the same rotational velocity.

\section{Summary}\label{sec:summary}
We present a catalogue of our LSBG sample 
which is selected from the $\alpha$.40-SDSS DR7 survey combination
and extends the parameter space of the previous LSBG samples.
This sample is an H{\sc{i}}-selected sample 
which is composed of 1129 LSBGs of various types of luminosity and morphology.

Based on the width of the H{\sc{i}} line and the absolute magnitude,
we investigate the Tully-Fisher relations (TFr) for a subsample from our
entire LSBG sample in the optical $B$, $g$ and $r$ and NIR $J$, $H$ and $K$ bands.
The TFrs in the NIR bands are tighter than those in the optical bands for our LSBG subsample.
This is because NIR magnitudes are less affected by internal attenuation and
recent episodes of star formation than optical bands.
In the optical bands, the LSBG subsample follows the fundamental TFr
which was defined for the normal spiral galaxies.
In the NIR bands, the slopes of the TFr lines for our LSBG subsample
are slightly different from those for the bright and nearby galaxies,
but are generally consistent with the TFrs for the bright and nearby galaxies
within the uncertainties. This conclusions agree with those from the previous TFr studies for LSBGs
\citep{Sprayberry1995,Zwaan1995,Verheijen1997}.

On the basis of the optical TFr results for our subsample,
we further estimated the mass-to-light ratio, disk scale length
 and mass surface density of our LSBG subsample. 
 Compared with the normal spiral galaxies with the same H{\sc{i}} line width. 
 the LSBGs have higher mass-to-light ratio, higher disk scale length, and 
 lower mass surface brightness than the HSB spiral galaxies,
 implying that LSBGs are mostly dominated by dark matter and
 they have low star formation at present.



\acknowledgments

We appreciate the anonymous referee for his/her constructive comments,
which helped us strengthen this paper.
We would like to gratefully thank Prof. Albert Bosma from Aix Marseille University, France,
who had face-to-face discussions with us to give us very helpful and suggestive
comments on the Tully-Fisher relation part of this paper.
We thank Dr. Tao Hong from National Astronomical Observatories, CAS for
helpful and constructive discussions and comments.
We thank Dr. Anastasis Ponomareva from Australian National University who had
private communications, shared her method of intrinsic extinction 
correction and also PhD thesis with us which is very helpful for us.
We also thank Prof. Galaz Gaspar from Pontificia Universidad Catolica de Chile,
who also gave us suggestive comments to help us improve this paper.
 
DW and WH is supported by the National Natural Science Foundation of China (NSFC)
 grant Nos.11733006 and 11403037, and the National Key R$\&$D Program of China 
 grant No. 2017YFA0402704. 
 CC is supported by the NSFC grant No. 11803044
 and is also supported in part by the Chinese Academy of Sciences (CAS), 
 through a grant to the CAS South America Center for Astronomy (CASSACA) in Santiago, Chile.
 ZM is supported by the the National Key R$\&$D Program of China grant No. 2017YFA0402600,
 NSFC grant No. U1531246 and the Major Applied Basic Research Program
of Guizhou Province (No. JZ[2014]2001-05). 
WY is supported by the NSFC grant Nos. 11390372
and 11773034. Both DW and CC are also supported by the Young Researcher Grant 
of National Astronomical Observatories, Chinese Academy of Sciences (NAOC) and
the Key Laboratory of Optical Astronomy, NAOC
%

\vspace{5mm}
\facilities{}


\software{
         }



\appendix

\begin{longrotatetable}
\begin{deluxetable*}{llllllllllllllll}
\tablecaption{A subset of the catalogue of our LSBG sample from the $\alpha$.40--SDSS DR7.\label{tab:cata1}}
\tablewidth{750pt}
\tabletypesize{\scriptsize}
\tablehead{
\colhead{AGC} &\colhead{W50}  &
\colhead{g} & \colhead{r} &\colhead{B} & \colhead{b/a} &\colhead{h(g)} &\colhead{R$_{eff}$(g)} &
\colhead{$\mu_{0,g}$} & \colhead{$\mu_{0,B}$} & \colhead{$\mu_{0,B}$} & \colhead{$\mu_{0,B}$} & \colhead{$\mu_{0,B}$} & 
\colhead{$\mu_{0,B}$} & 
 \colhead{T1} &\colhead{T2} \\ 
\colhead{ } &\colhead{(km s$^{-1}$)} &
\colhead{(mag)} &\colhead{(mag)} &\colhead{(mag)} &\colhead{ } &\colhead{(kpc)} &\colhead{(kpc)} &
\colhead{(AUTO)} &\colhead{( AUTO)} &\colhead{ISO} &\colhead{ISOCOR} & \colhead{PETRO} & \colhead{AVE }&
\colhead{ } & \colhead{ }
} 
\startdata
7200 &     77$\pm$      2 &14.45$\pm$0.003 &14.10$\pm$0.003 &14.78$\pm$0.005 &0.62 &1.42 &1.80 &22.17 &22.53 &23.32 &22.86 &22.49 &22.80 &S &D
 \\
7223 &     80$\pm$      1 &14.87$\pm$0.004 &14.48$\pm$0.005 &15.22$\pm$0.007 &0.90 &1.09 &1.79 &22.39 &22.84 &23.96 &23.32 &22.81 &23.23 &S &D
 \\
7236 &     55$\pm$      2 &17.01$\pm$0.010 &16.72$\pm$0.012 &17.32$\pm$0.016 &0.91 &0.70 &0.65 &24.38 &24.63 &26.96 &26.39 &24.51 &25.62 &L &D
 \\
7239 &    111$\pm$      2 &13.58$\pm$0.003 &13.20$\pm$0.003 &13.92$\pm$0.004 &0.84 &2.10 &3.00 &22.47 &22.86 &23.74 &23.12 &22.83 &23.14 &S &M
 \\
7254 &     96$\pm$      1 &14.39$\pm$0.004 &14.00$\pm$0.004 &14.74$\pm$0.006 &0.70 &1.85 &2.28 &22.80 &23.15 &24.13 &23.35 &23.10 &23.43 &S &D
 \\
7307 &     52$\pm$      0 &16.52$\pm$0.010 &16.19$\pm$0.011 &16.84$\pm$0.015 &0.36 &1.93 &1.31 &24.29 &24.59 &27.39 &25.88 &24.61 &25.62 &L &D
 \\
7424 &     38$\pm$      1 &14.61$\pm$0.004 &14.15$\pm$0.004 &15.00$\pm$0.007 &0.79 &1.33 &1.73 &22.48 &22.89 &23.60 &22.87 &22.80 &23.04 &S &D
 \\
7547 &     71$\pm$      1 &15.18$\pm$0.008 &14.89$\pm$0.008 &15.49$\pm$0.013 &0.60 &1.78 &2.04 &23.33 &23.29 &24.99 &24.54 &23.23 &24.01 &S &D
 \\
\enddata
\end{deluxetable*}
\end{longrotatetable}


\begin{thebibliography}{}

\bibitem[\protect\citeauthoryear{Aaronson \& Mould.}{1983}]{Aaronson1983}Aaronson, M. \&  Mould, J, 1983, \apj, 265, 1
\bibitem[\protect\citeauthoryear{Abazajian et al.}{2009}]{Abazajian2009}Abazajian, K. N., Adelman-McCarthy, J. K., Agüeros, M. A., et al., 2009, \apjs, 182, 543
\bibitem[\protect\citeauthoryear{Adelman-McCarthy et al.}{2006}]{Adelman-McCarthy2006}Adelman-McCarthy, J. K., Agüeros, M. A., Allam, S. S., et al., 2006, \apjs, 162, 38A
\bibitem[\protect\citeauthoryear{Bertin \& Arnouts}{1996}]{Bertin1996} Bertin, E., \& Arnouts, S., 1996, \aaps, 117, 393 
\bibitem[\protect\citeauthoryear{Begum et al.}{2008}]{Begum2008}Begum, A., Chengalur, J. N., Karachentsev, I. D., \& Sharina, M. E., 2008, \mnras, 386, 138
\bibitem[\protect\citeauthoryear{Bell $\&$ de Jong}{2001}]{Bell2001}Bell, E. F. \& de Jong, R. S., 2001, \apj, 550, 212
\bibitem[\protect\citeauthoryear{Bergvall et al.}{1999}]{Bergvall1999}Bergvall, N., Ronnback, J., Masegosa, J., \& Ostlin, G., 1999, A$\&$A, 341, 697
\bibitem[\protect\citeauthoryear{Boissier et al.}{2016}]{Boissier2016}Boissier, S., Boselli, A., Ferrarese L., et al., 2016, A$\&$A, 593, 126
\bibitem[\protect\citeauthoryear{Bothun}{1982}]{Bothun1982}Bothun, G. D., 1982, \apjs, 50, 39
\bibitem[\protect\citeauthoryear{Bothun et al.}{1987}]{Bothun1987}Bothun, G. D., Impey, D. C., Malin, F. D. \& Mould, R. J., 1987, \aj, 94, 1
\bibitem[\protect\citeauthoryear{Bothun et al.}{1990}]{Bothun1990} Bothun, G. D., Schombert, J. M., Impey, C. D. \& Schneider, S. E., 1990, \apj, 360, 427
\bibitem[\protect\citeauthoryear{Bothun et al.}{1997}]{Bothun1997} Bothun G. D. et al., 1997, \pasp, 109, 745
\bibitem[\protect\citeauthoryear{Bottinelli et al.}{1986}]{Bottinelli1986}Bottinelli, L., Gouguenheim, L., Paturel, G., \& Teerikorpi, P., 1986, A$\&$A, 156, 157
\bibitem[\protect\citeauthoryear{Broeils}{1992}]{Broeils1992} Broeils A. H., 1992, PhD thesis, Univ. Groningen
\bibitem[\protect\citeauthoryear{Burkholder et al.}{2001}]{Burkholder2001} Burkholder, V., Impey, C. \& Sprayberry, D., 2001, \aj, 122, 2318
\bibitem[\protect\citeauthoryear{Burstein}{1982}]{Burstein1982}Burstein, D., 1982, \apj, 253, 539
\bibitem[\protect\citeauthoryear{Ceccarelli et al.}{2012}]{Ceccarelli2012}Ceccarelli L., Herrera-Camus R., Lambas D. G., Galaz G., Padilla N. D., 2012, \mnras, 426, 6
\bibitem[\protect\citeauthoryear{Chung et al.}{2002}]{Chung2002} Chung, A., van Gorkom, J. H., O'Neil, K., \& Bothun, D. G., 2002, \aj, 123, 2387
\bibitem[\protect\citeauthoryear{Courteau et al.}{1997}]{Courteau1997}Courteau, S., 1997, \aj, 114, 2402
\bibitem[\protect\citeauthoryear{Courteau et al.}{2007}]{Courteau2007}Courteau, S., Dutton, A. A., van den Bosch, F. C, et al. 2007, \apj, 671, 203
\bibitem[\protect\citeauthoryear{Courtois et al.}{2009}]{Courtois2009}Courtois, M. H., Tully, B. R., Fisher, R. J., Bonhomme, N., \& Zavodny, M. 2009, \aj, 138, 1938
\bibitem[\protect\citeauthoryear{Das et al.}{2009}]{Das2009} Das M. et al., 2009, \apj, 693, 1300
\bibitem[\protect\citeauthoryear{de Blok et al.}{1995}]{de Blok1995}de Blok, W. J. G., van der Hulst, J. M., \& Bothun, G. D., 1995, \mnras, 274, 235
\bibitem[\protect\citeauthoryear{de Blok et al.}{1996}]{de Blok1996}de Blok, W. J. G., McGaugh, S. S., \& van der Hulst, J. M., 1996, \mnras, 283, 18
\bibitem[\protect\citeauthoryear{de Vaucouleurs}{1959}]{de Vaucouleurs1959}de Vaucouleurs G., 1959, Handbuch der Physik, 53, 311
\bibitem[\protect\citeauthoryear{de Vaucouleurs} {1974}]{de Vaucouleurs1974}de Vaucouleurs G.,1974, IAU Symp., 58, 1
\bibitem[\protect\citeauthoryear{de Vaucouleurs} {1975}]{de Vaucouleurs1975}de Vaucouleurs G.,1975, \apjs, 29, 193
\bibitem[\protect\citeauthoryear{de Vaucouleurs et al.} {1976}]{de Vaucouleurs1976}de Vaucouleurs G., de Vaucouleurs A., \& Corwin H. G., 1976, Second Reference Catalogue of Bright Galaxies. University of Texas Press, Austin
\bibitem[\protect\citeauthoryear{Dobrycheva et al.}{2017}]{Dobrycheva2017} Dobrycheva et al., 2017, arXiv:1712.08955
\bibitem[\protect\citeauthoryear{Driver et al.}{2011}]{Driver2011}Driver, S. P., et al., 2011, \mnras, 413, 971
\bibitem[\protect\citeauthoryear{Du et al.}{2015}]{Du2015} Du, W., Wu, H., Lam, Man~I, et al., 2015, \aj, 149, 199
\bibitem[\protect\citeauthoryear{Dunn}{2010}]{Dunn2010}Dunn M. J., 2010, \mnras, 408, 392
\bibitem[\protect\citeauthoryear{Fouque et al.} {1990}]{Fouque1990}Fouque, P., Bottinelli, L., \& Gouguenheim, L., 1990, \apj, 349, 1
\bibitem[\protect\citeauthoryear{Freedman}{1990}]{Freedman1990}Freedman, L. W., 1990, \apj, 355, L35
\bibitem[\protect\citeauthoryear{Galaz et al.}{2002}]{Galaz2002} Galaz, G., Dalcanton, J. J., Infante, L., \& Treister, E., 2002, \aj, 124, 1360
\bibitem[\protect\citeauthoryear{Galaz et al.}{2011}]{Galaz2011} Galaz G. et al., 2011, \apj, 728, 74
\bibitem[\protect\citeauthoryear{Galaz et al.}{2015}]{Galaz2015}Galaz, G., Milovic, C., Suc, V. et al. 2015, ApJL, 815, 29
\bibitem[\protect\citeauthoryear{Giovanelli et al.} {1997}]{Giovanelli1997} Giovanelli, R., Haynes, M. P., Herter, T. et al., 1997, \aj, 113, 53
\bibitem[\protect\citeauthoryear{Giovanelli et al.} {2005}]{Giovanelli2005} Giovanelli, R., Haynes, M. P., Kent, B. R., et al., 2005, \aj, 130, 2598
\bibitem[\protect\citeauthoryear{Gurovich et al.} {2010}]{Gurovich2010}Gurovich, S., Freeman, K., Jerjen, H., Staveley-Smith, L., \& Puerari, I., 2010, \aj, 140, 663
\bibitem[\protect\citeauthoryear{Haberzettl et al.}{2007}]{Haberzettl2007} Haberzettl L. et al., 2007, A$\&$A, 471, 787
\bibitem[\protect\citeauthoryear{He et al.}{2013}]{He2013} He, Y. Q., Xia, X. Y., Hao, C. N., et al., 2013, \apj, 773, 37
\bibitem[\protect\citeauthoryear{Haynes et al.} {2011}]{Haynes2011} Haynes, M. P., Giovanelli, R., Martin, A. M. et al., 2011, \aj, 142, 170
\bibitem[\protect\citeauthoryear{Haynes et al.} {2018}]{Haynes2018} Haynes, M. P., Giovanelli, R., Kent, B. R. et al., 2018, \apj, 861, 49
\bibitem[\protect\citeauthoryear{Ho}{2007}]{Ho2007} Ho, Luis, 2007, \apj, 668, 94
\bibitem[\protect\citeauthoryear{Hubble} {1926}]{Hubble1926} Hubble, E. \ 1926, \apj, 64, 321
\bibitem[\protect\citeauthoryear{Impey, Bothun \& Malin}{1988}]{Impey1988}Impey, C. D., Bothun, G. D. \& Malin, F. D., 1988, \apj, 330, 634
\bibitem[\protect\citeauthoryear{Impey et al.}{1996}]{Impey1996}Impey, C. D., Sprayberry, D., Irwin, M. J., \& Bothun, G. D., 1996, \apjs, 105, 209
\bibitem[\protect\citeauthoryear{Impey \& Bothun}{1997}]{Impey1997} Impey, C. \& Bothun, G., 1997, AR\&AA, 35, 267
\bibitem[\protect\citeauthoryear{Impey et al.}{2001}]{Impey2001} Impey C., Burkholder V., Sprayberry D., 2001, \aj, 122, 2341
\bibitem[\protect\citeauthoryear{Infante}{1987}]{Infante1987} Infante L 1987, A$\&$A, 183, 177
\bibitem[\protect\citeauthoryear{Knezek} {1993}]{Knezek1993} Knezek P. M., 1993, PhD thesis, Univ. Massachusetts
\bibitem[\protect\citeauthoryear{Kniazev et al.}{2004}]{Kniazev2004} Kniazev, A. Y., Grebel, E. K., Pustilnik, S. A. et al., 2004, \aj, 127, 704
\bibitem[\protect\citeauthoryear{Kron}{1980}]{Kron1980} Kron R. G. 1980, \apj, 43, 305
\bibitem[\protect\citeauthoryear{Lei et al.}{2018}]{Lei2018}Lei F. J., Wu H., Du W., et al., 2018, \apjs, 235, 18
\bibitem[\protect\citeauthoryear{Liske et al.}{2015}]{Liske2015}Liske, J., et al., 2015, \mnras, 452, 2087
\bibitem[\protect\citeauthoryear{Lisker et al.}{(2007}]{Lisker2007}Lisker, T., Grebel, E. K., Binggeli, B., \& Katharina, G., 2007, \apj, 660, 1186
\bibitem[\protect\citeauthoryear{Liu et al.}{2008}]{Liu2008}Liu, F. S., Xia, X. Y., Mao, S., Wu, H., \& Deng, Z. G., 2008, \mnras, 385, 23
\bibitem[\protect\citeauthoryear{Malhotra et al. }{1996}]{Malhotra1996}Malhotra, S., Spergel, D. N., Rhoads, J. E., \& Li, J., 1996, \apj, 473, 687
\bibitem[\protect\citeauthoryear{Masters et al.}{2008}]{Masters2008} Masters et al., 2008, \aj, 135, 1738
\bibitem[\protect\citeauthoryear{Matthews et al.}{1998}]{Matthews1998}Matthews, L. D., van Driel, W., \& Gallagher, J. S., 1998, \aj, 116, 2196
\bibitem[\protect\citeauthoryear{Matthews et al.}{2001}]{Matthews2001}Matthews, L. D., van Driel, W., Monnier-Ragaigne, D., 2001, A$\&$A, 365, 1
\bibitem[\protect\citeauthoryear{McGaugh \& Bothun}{1994}]{McGaugh1994}McGaugh, S. S., \& Bothun, G. D., 1994, \aj, 107, 530
\bibitem[\protect\citeauthoryear{McGaugh et al.}{1995}]{McGaugh1995}McGaugh, S. S., Bothun, G. D., \& Schombert, J. M., 1995, \aj, 110, 573
\bibitem[\protect\citeauthoryear{McGaugh}{1996}]{McGaugh1996} McGaugh S. S., 1996, \mnras, 280, 337
\bibitem[\protect\citeauthoryear{McGaugh et al.}{2000}]{McGaugh2000}McGaugh, S. S., Schombert, J. M., Bothun, G. D., \& de Blok, W. J. G., 2000, \apj, 533, 99
\bibitem[\protect\citeauthoryear{McGaugh}{2005}]{McGaugh2005}McGaugh, S. S., 2005, \apj, 632, 859
\bibitem[\protect\citeauthoryear{McGaugh}{2012}]{McGaugh2012}McGaugh, S. S., 2012, \aj, 143, 40
\bibitem[\protect\citeauthoryear{Minchin et al.}{2004}]{Minchin2004} Minchin, R. F. et al., 2004, \mnras, 355, 1303
\bibitem[\protect\citeauthoryear{Nair \& Abraham}{2010}]{Nair2010}Nair, B. P. \& Abraham, G. R., 2010, \apjs, 186, 427
\bibitem[\protect\citeauthoryear{O'Neil et al.}{2000}]{O'Neil2000}O'Neil, K., Bothun, G. D., \& Schombert J.,  2000, \aj, 119, 136
\bibitem[\protect\citeauthoryear{O'Neil et al.}{2004}]{O'Neil2004}O'Neil, K., Bothun, G. D., van Driel, W., \& Monnier Ragaigne, D.,  2004, A$\&$A, 428, 823
\bibitem[\protect\citeauthoryear{Papastergis et al.}{2016}]{Papastergis2016}Papastergis, E., Adams, E. A. K., \& van der Hulst, J. M., 2016, 593, 39
\bibitem[\protect\citeauthoryear{Pizagno et al. }{2007}]{Pizagno2007}Pizagno, J., Prada, F., Weinberg, D. H. et al., 2007, \aj, 134, 945
\bibitem[\protect\citeauthoryear{Peng et al.}{2002}]{Peng2002}Peng, C. Y., Ho, L. C., Impey, C. D., \& Rix, H., 2002, \aj, 124, 266
\bibitem[\protect\citeauthoryear{Pohlen \& Trujillo}{2006}]{Pohlen2006}Pohlen, M. \& Trujillo, I., 2006, A$\&$A, 454, 759 
\bibitem[\protect\citeauthoryear{Ponomareva et al.}{2017}]{Ponomareva2017}Ponomareva et al., 2017, \mnras, 469, 2387
\bibitem[\protect\citeauthoryear{Ponomareva et al.}{2017}]{Ponomareva2017b}Ponomareva, A. 2017, Understanding disk galaxies with the Tully-Fisher relation [Groningen]: Rijksuniversiteit Groningen
\bibitem[\protect\citeauthoryear{Sabatini et al.}{2003}]{Sabatini2003}Sabatini S., Roberts S., \& Davies J., 2003, Ap$\&$SS, 285, 97
\bibitem[\protect\citeauthoryear{Sandage \& Binggeli}{1984}]{Sandage1984} Sandage, a. \& Binggeli, B., 1984, \aj, 89, 919
\bibitem[\protect\citeauthoryear{Sakai et al.}{2000}]{Sakai2000}Sakai, S., Mould, J. R., Hughes, S. M. G., et al., 2000, \apj, 529, 698
\bibitem[\protect\citeauthoryear{Schlafly \& Finkbeiner}{2011}]{Schlafly2011} Schlafly, E. F. \& Finkbeiner, D. P., 2011, \apj, 737, 103
\bibitem[\protect\citeauthoryear{Schlegel et al.}{1998}]{Schlegel1998}Schlegel D., Finkbeiner D., \& Davis M., 1998, \apj, 500, 525
\bibitem[\protect\citeauthoryear{Schombert et al.}{1992}]{Schombert1992} Schombert J. et al., 1992, \aj, 103, 1107
\bibitem[\protect\citeauthoryear{Schombert et al.}{2013}]{Schombert2013} Schombert J. et al., 2013, \aj, 146, 41
\bibitem[\protect\citeauthoryear{Shao et al.}{2007}]{Shao2007}Shao, Z. Y., Xiao, Q. B., Shen S. Y. et al.,2007, \apj, 659, 1159 
\bibitem[\protect\citeauthoryear{Simons et al.}{2015}]{Simons2015} Simons R. C., Kassin S. A., Weiner B. J. et al., 2015, \mnras, 452, 986
\bibitem[\protect\citeauthoryear{Smith et al.}{2002}]{Smith2002}Smith, J. A., Tucker, D. L., Kent, S., et al., 2002, \aj, 123, 2121
\bibitem[\protect\citeauthoryear{Sorce et al.} {2012}] {Sorce2012}Sorce, J. G., Tully, R. B., Courtois, H. M., 2012, \apj, 758, 12
\bibitem[\protect\citeauthoryear{Sorce et al.}{2012}]{Sorce2012a}Sorce, J. G., Courtois, H. M., \& Tully, R. B., 2012, \aj, 144, 133
\bibitem[\protect\citeauthoryear{Sorce et al.}{2013}]{Sorce2013}Sorce, J. G., Courtois, H. M., Tully, R. B., et al., 2013, \apj, 765, 94
\bibitem[\protect\citeauthoryear{Sprayberry et al.}{1993}]{Sprayberry1993}Sprayberry, D., Impey, D. C., Irwin, M. J., McMahon, R. G. \& Bothun, D. G., 1993, \apj, 417, 114
\bibitem[\protect\citeauthoryear{Sprayberry, Bernstein \& Impey}{1995}]{Sprayberry1995}Sprayberry, D., Bernstein, M. G. \& Impey, D. C., 1995, \apj, 438, 72
\bibitem[\protect\citeauthoryear{Theureau et al. }{2007}]{Theureau2007}Theureau, G., Hanski, M. O., Coudreau, N., Hallet, N., \& Martin, J. M., 2007, A$\&$A, 465, 71
\bibitem[\protect\citeauthoryear{Tiley et al.}{2016}]{Tiley2016}Tiley, A. L., Bureau, M., Saintonge, A. et al., 2016, \mnras, 461, 3494
\bibitem[\protect\citeauthoryear{Trachternach et al.}{2006}]{Trachternach2006} Trachternach, C., Bomans, D. J., Haberzettl, L., \& Dettmar, R. J., 2006, A$\&$A, 458, 341
\bibitem[\protect\citeauthoryear{Trachternach et al.} {2009}]{Trachternach2009}Trachternach, C., de Blok, W. J. G., McGaugh, S. S., van der Hulst, J. M., \& Dettmar, R. -J., 2009, A$\&$A, 505, 577
\bibitem[\protect\citeauthoryear{Tully \& Fisher}{1977}]{Tully1977}Tully R. B. \& Fisher J. R., 1977, A$\&$A, 54, 661
\bibitem[\protect\citeauthoryear{Tully \& Fouque}{1985}]{Tully1985}Tully R. B. \& Fouque P., 1985, \apjs, 58, 67
\bibitem[\protect\citeauthoryear{Tully et al.}{1998}]{Tully1998} Tully, R. B., Pierce, M. J., Huang, J. S. et al.,1998, \aj, 115, 2264
\bibitem[\protect\citeauthoryear{van der Hulst et al.}{1993}]{van der Hulst1993} van der Hulst J. M. et al., 1993, \aj, 106, 548
\bibitem[\protect\citeauthoryear{van Dokkum et al.}{2015}]{van Dokkum2015} van Dokkum P., Abraham R., Merritt A. et al., 2015, \apjl, 798, L45
\bibitem[\protect\citeauthoryear{Verheijen}{1997}]{Verheijen1997} Verheijen, M. A., 1997, PhD thesis, Univ. Groningen 
\bibitem[\protect\citeauthoryear{Williams et al.}{2016}]{Williams2016} Williams, R. P., Baldry, I. K., Kevin, L. S., et al., 2016, \mnras, 463, 2746
\bibitem[\protect\citeauthoryear{Willick}{1994}]{Willick1994} Willick, 1994, \apjs, 92, 1
\bibitem[\protect\citeauthoryear{Wu et al.}{2002}]{Wu2002} Wu, H., Burstein, D., Deng, Z. G., et al., 2002, \aj, 123, 1364
\bibitem[\protect\citeauthoryear{Zaritsky et al.}{2014}]{Zaritsky2014}Zaritsky, D., Courtois, H., Munoz-Mateos, J. -C., et al., 2014, \aj, 147, 134
\bibitem[\protect\citeauthoryear{Zheng et al.}{1999}]{Zheng1999} Zheng, Z. Y., Shang, Z. H., Su, H. J., et al., 1999, \aj, 117, 2757
\bibitem[\protect\citeauthoryear{Zhong et al.}{2008}]{Zhong2008} Zhong, G. H., Liang, Y. C., Liu, F. S., et al., 2008, \mnras, 391, 986
\bibitem[\protect\citeauthoryear{Zwaan et al.}{1995}]{Zwaan1995}Zwaan, M., van der Hulst, J., de Blok, W., \& McGaugh, S., 1995, \mnras, 273, L35

\end{thebibliography}
\end{document}